\pdfoutput=1

\documentclass[11pt]{article}

\usepackage[preprint]{acl}

\usepackage{times}
\usepackage{latexsym}
\usepackage{hyperref}

\usepackage[T1]{fontenc}

\usepackage[utf8]{inputenc}

\usepackage{microtype}

\usepackage{inconsolata}

\usepackage{graphicx}

\usepackage{dirtytalk}
\usepackage{booktabs}
\usepackage{enumitem}
\usepackage{float}
\usepackage{amsmath}
\usepackage{multirow}

\usepackage{tcolorbox}
\usepackage{float}
\usepackage{newfloat}
\usepackage[skip=2pt]{caption}
\tcbuselibrary{breakable}

\usepackage{fontawesome5}

\DeclareFloatingEnvironment[
    fileext=lob,
    listname=List of Text Boxes,
    name=Prompt,                       
    placement=htbp,
]{textbox}

\definecolor{lightgreen}{HTML}{e8f4ea}
\definecolor{darkgreen}{HTML}{b8d8be}

\DeclareFloatingEnvironment[
    fileext=lob,
    listname=List of Text Boxes,
    name=Codebook,                       
    placement=htbp,
]{cb_textbox}

\definecolor{lightblue}{HTML}{b7c6d4}
\definecolor{darkblue}{HTML}{02075D}

\usepackage{tabularx}

\title{How Do We Engage with Other Disciplines? A Framework to Study Meaningful Interdisciplinary Discourse in Scholarly Publications}

\author{
   Bagyasree Sudharsan \,\,
   \textbf{Alexandria Leto} \,\,
   \textbf{Maria Leonor Pacheco} \\
   University of Colorado Boulder \\
   \texttt{\{bagyasree.sudharsan, alexandria.leto, maria.pacheco\}@colorado.edu} \\
}

\begin{document}
\maketitle
\begin{abstract}

With the rising popularity of interdisciplinary work, there is a growing need to understand how publications incorporate ideas from multiple disciplines. Existing computational approaches such as affiliation diversity, keyword matching, and citation patterns do not account for \textit{how} individual citations are used to advance the citing work. Prior studies have proposed taxonomies to classify citation purpose. However, these frameworks are not well-suited to interdisciplinary research and do not provide general measures of engagement \textit{depth}. To address these limitations, we propose a framework for evaluating citation engagement in scholarly publications. We introduce a citation purpose taxonomy tailored to interdisciplinary work, supported by an annotation study and a model for automatically predicting a citation's purpose. We then leverage signal from a citation's purpose and the sections in which it appears to predict a single measure of how deeply a publication engages with a citation. We demonstrate the utility of our framework through a large-scale analysis of publications at the intersection of NLP and Computational Social Science.

\end{abstract}

\section{Introduction}
Interdisciplinary research involves the integration of concepts, perspectives, theories, methods, and data from two or more bodies of specialized knowledge to construct creative solutions \cite{NAS2005Interdisciplinary}. This definition highlights knowledge integration as a central component of interdisciplinarity, suggesting that it should be assessed not only in terms of the amount of knowledge integrated but also how it is combined. Meta-analyses which seek to understand the convergence of, and interplay between, different fields should therefore aim to capture the \textit{depth} with which an article engages with the work it cites. 


Contrary to this, much of the existing meta-analytic work has focused on broad, scalable indicators (e.g., diversity and entropy of citations) 
~\cite{porter_is_2009,noorden_interdisciplinary_2015, cassi2017, qun2023} that emphasize the quantity and diversity of knowledge sources, but offer limited insight into how deeply or substantively those sources are integrated. 
For example, an article may cite a broad set of disciplines in its \textit{Introduction} or \textit{Related Work} sections without explicitly building on the cited work's ideas or using them to inform methodological choices. 


\begin{figure}
    \centering
    \includegraphics[width=\linewidth]{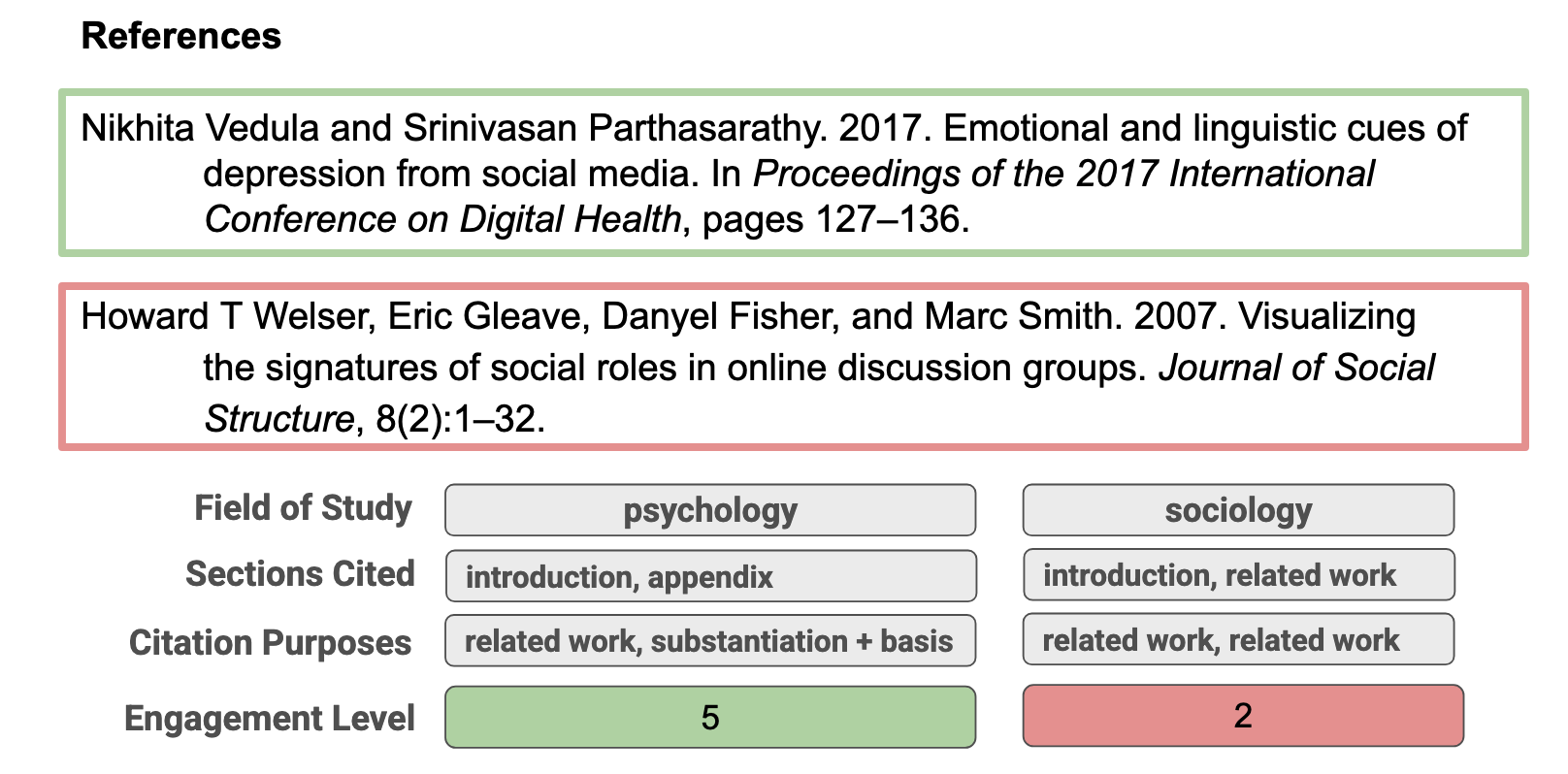}
    \caption{Framework for evaluating a publication's level of engagement with citing work based on the section and purpose of all in-text references.}
    \label{fig:tr_fig}
\end{figure}

Motivated by the need to go beyond traditional measures of interdisciplinarity, we propose a framework for characterizing the depth with which an article engages with the works it cites.\footnote{\faGithub ~\href{https://github.com/blast-cu/interdisciplinary-engagement}{Code, models, and data}.}
We build on the framework by \citet{leto-etal-2024-first}, which is motivated by the intuition that the section in which a citation appears signals different levels of engagement. It extracts in-text citations from interdisciplinary papers and maps them to the section in which they occur and identifies whether they are out-of-discipline from the citing work. We extend this framework by introducing a novel taxonomy to represent the purpose for which an author includes a given citation. Although numerous citation-purpose classification schemes have been proposed ~\cite{jurgens-etal-2018-measuring, abu-jbara-etal-2013-purpose, lauscher-etal-2022-multicite, cohan-etal-2019-structural, teufel-etal-2006-automatic, pride-and-knoth-authoritative}, existing taxonomies are insufficient for analyzing interdisciplinary citations, either because they lack relevant categories, include categories that are ill-suited to cross-disciplinary contexts, or fail to incorporate adequate contextual information. 

Our taxonomy, shown in Figure \ref{fig:tr_fig}, addresses these limitations by situating each citation within its local argumentative context and characterizing citation purpose through categories designed to reflect the different forms of interdisciplinary engagement that arise when authors draw on work beyond their primary field. The taxonomy was developed from the ground up through an inductive annotation study, allowing categories to emerge from close analysis of real citation practices rather than being imposed a priori. 

We operationalize our framework, integrating signal from a citation's purpose and the section in which it appears into a single \textbf{Citation Engagement Predictor (CEP)}. Our CEP provides a measure of the depth of a publication's engagement with a work it cites, allowing practitioners to gauge engagement at a glance. We ground our proposed CEP in an annotation study and statistical analysis. These analyses find that high-engagement citations are rare and some purposes are more common in interdisciplinary works. Together with our qualitative analysis, they suggest that a dedicated effort towards making citations more integrative would result in better interdisciplinary work.  

Our framework, annotation studies, and analysis are developed in the context of Natural Language Processing (NLP) research, positioning these contributions as a first step toward a broader framework for studying interdisciplinary engagement across fields. We focus specifically on work at the intersection of NLP and Computational Social Science (CSS). Although research in this area has grown substantially over the past decade \cite{Grimmer_Stewart_2013}, the quality of NLP+CSS publications has been questioned, with critiques arguing that such work often falls short of social science standards of rigor \cite{christian_baden_three_2022} and makes limited substantive contributions to those fields \cite{mccarthy-dore-2023-theory}. We use our framework to further examine these concerns and to evaluate the extent to which citation engagement can be reliably studied at scale, including through the use of automated methods. We make the following contributions:

\begin{enumerate}
    \item A novel annotation framework designed to capture citation purpose in interdisciplinary NLP work. We accompany this with a dataset of 950 annotated in-text citations from NLP+CSS papers and an automatic classifier for predicting a citation's purpose according to our taxonomy.

    \item A Citation Engagement Predictor designed to provide a rating of a publication's engagement with a given cited work based on signal from the purpose and section of corresponding in-text references.
    
    \item A case study in NLP+CSS work demonstrating the usefulness of our framework. We find that articles with many high-engagement interdisciplinary citations are more principled and better motivated, while many within-discipline citations provide more rigorous dataset analysis. Interdisciplinary citations are more often engaged with at the idea- and claim-level, while within-discipline citations drive methodology.
    
\end{enumerate}

\section{Related Work}

    


Numerous studies have introduced frameworks for classifying citation purpose or intent, but these approaches are not designed to capture how authors engage with references from outside their primary discipline. For example, \citet{jurgens-etal-2018-measuring} and \citet{pride-and-knoth-authoritative} introduce datasets that are heavily dominated by a \textit{Background} category designed to denote when a reference provides relevant information for the domain of the article. By treating all background citations as equivalent, this definition fails to distinguish between references that provide foundational knowledge necessary for interpreting the work and those that merely provide general or loosely related context. Similarly, their \textit{Extension} category is narrowly defined to apply only when authors directly extend a method or dataset; a scenario that is relatively rare in interdisciplinary contexts, where engagement often takes other forms. In addition, their framework relies on a fixed citation context of approximately 300 surrounding characters, which may be insufficient to capture the broader argumentative role needed to accurately determine citation purpose. \citet{lauscher-etal-2022-multicite} and \citet{abu-jbara-etal-2013-purpose} address this limitation by considering a window of sentences surrounding each citation — an approach that we adopt in this work.

\citet{teufel-etal-2006-automatic} propose a highly fine-grained scheme for classifying citation function, with multiple sub-categories that distinguish forms of comparison and contrast. While this level of detail can be valuable, the resulting categories primarily capture evaluative stance and intra-discipline relations such as direct comparison and contrast. 
Additionally, their dataset labels over 60\% of citations as \textit{Neutral}, limiting its ability to distinguish meaningful forms of engagement. In contrast, our scheme places only $\sim$ 38.5\% of citations into similar general categories (see Table \ref{tab:ann-counts}), allowing for finer differentiation among types of interdisciplinary use. 

At the opposite side of the spectrum, \citet{cohan-etal-2019-structural} introduce a taxonomy with only three high-level categories, which are too coarse-grained to adequately capture variation in the depth of citation engagement required for our analysis.



\section{Data}
\label{sec:data}
In this section, we discuss our methods for collecting and processing our dataset of interdisciplinary NLP+CSS papers.

\subsection{Interdisciplinary Article Collection}

We construct a dataset of CSS papers from the ACL Anthology\footnote{\url{https://aclanthology.org/}} published between 2014-2025. To do so, we build on a pre-existing data collection methodology \cite{leto-etal-2024-first}, where an automatic track classifier is used to identify NLP+CSS publications. Approximately 1,800 abstracts were labeled with their track using conference programs from 2021-2023. Papers categorized under \say{Computational Social Science and Cultural Analytics} serve as positive examples, while negative examples are drawn from all other tracks. We augment the positive set with papers from the ACL Special Interest Group on Language Technologies for the Socio-Economic Sciences and Humanities (SIGHUM) workshop and the Natural Language Processing and Computational Social Science (NLP+CSS) workshop. The resulting dataset comprises 497 positive and 2,692 negative examples, for a total of 3,189 labeled abstracts. The distribution of labels over time is shown in Figure~\ref{fig:track_labels_year}. We use this labeled dataset to train a binary classifier to identify CSS papers. Specifically, we fine-tune RoBERTa-base \cite{gururangan-etal-2020-dont} on abstract text. Details are provided in Appendix~\ref{app:data-collect-details}.

We download all long publications in the ACL anthology from 2014-2025. Then, we use the classifier (details in Appendix \ref{app:track-classifier}) to identify 1,534 of the 17,258 unlabeled articles as NLP+CSS publications. Combined with the initial positively-labeled articles, this results in a dataset of 2,031 NLP+CSS articles.

\subsection{Citation Mapping}

After the CSS papers are identified, their content is processed using Grobid\footnote{\url{https://github.com/kermitt2/grobid}}, then converted into dictionary representation using SciPDF Parser\footnote{\url{https://github.com/titipata/scipdf_parser}}. The resulting dictionary contains a breakdown of the textual content, including the reference section. Section names are mapped to canonical sections \textit{Introduction}, \textit{Related Work}, \textit{Method}, \textit{Experiments}, \textit{Conclusion}, \textit{References}, and \textit{Appendix} (details in Appendix \ref{app:sec-mapping}). The in-text references are then extracted from each section and assigned to their entry in the \textit{References} section, connecting them to the title of the cited publication, the publication venue and other publication meta-data. Finally, cited works are mapped to a \say{field of study} by querying the Semantic Scholar API \cite{Kinney2023TheSS} or using a string-matching approach. All citations with a field of study outside of \say{Computer Science} and \say{Linguistics} are determined to be out-of-discipline. 

\begin{figure}
    \centering
    \includegraphics[width=\linewidth]{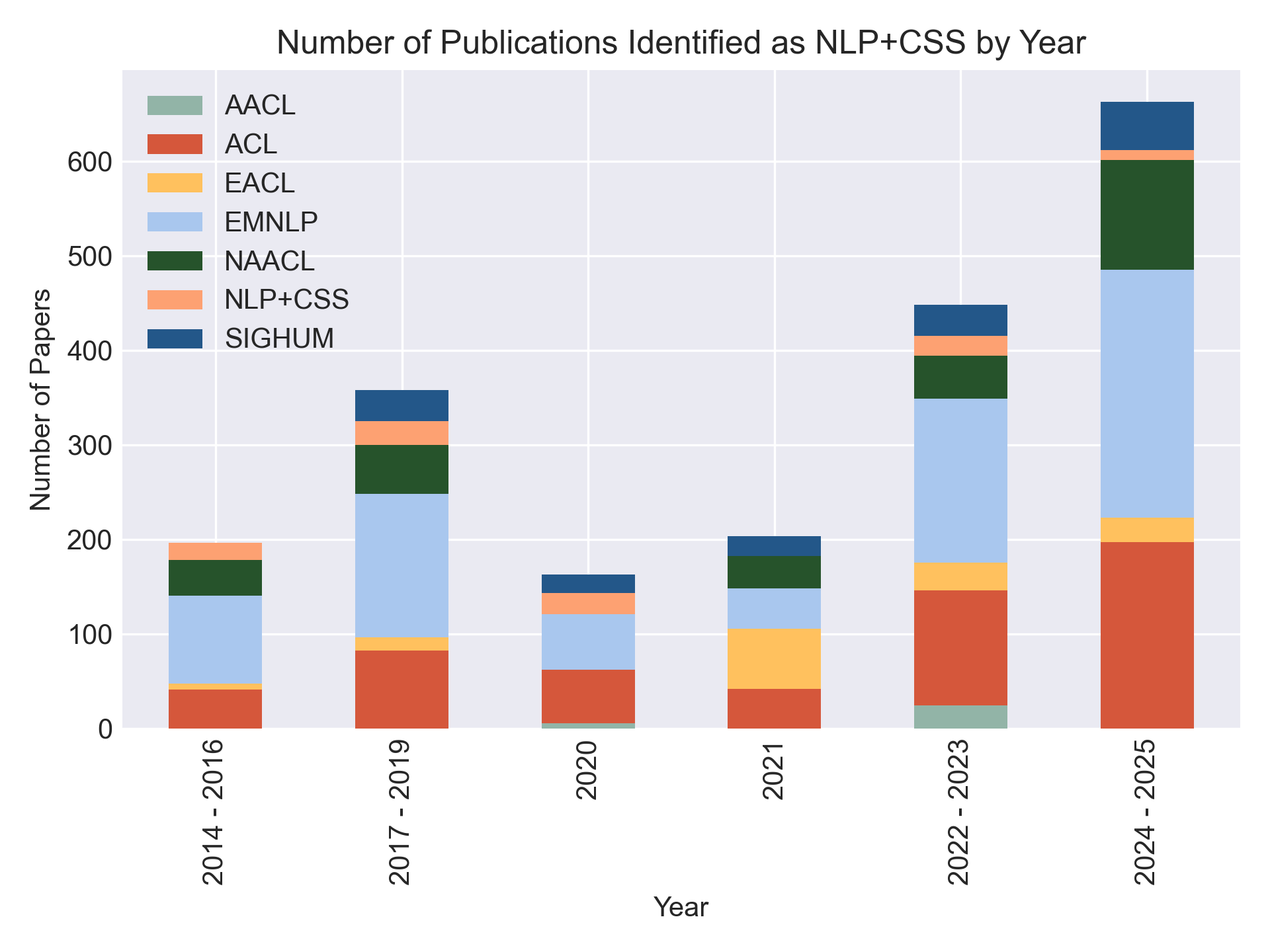}
    \caption{Dataset of NLP+CSS articles across years and venues}
    \label{fig:article_distribution}
\end{figure}

Our resulting dataset of NLP+CSS work contains 2,031 articles (see Figure \ref{fig:article_distribution}). From these, we extracted 50,050 in-text references to 19,051 unique cited works. 20,319 of these in-text references (17,996 cited works) were determined to be out-of-discipline, for an average of $\sim10$ out-of-discipline in-text citations per article.


\newcommand{\specialcell}[2][l]{%
  \begin{tabular}[#1]{@{}l@{}}#2\end{tabular}}

\begin{table*}[t]
    \centering
    \resizebox{\textwidth}{!}{%
    \begin{tabular}{ll}
    \toprule
        \textbf{Category} & \textbf{Explanation} \\
        \midrule
        \specialcell[c]{\textbf{SUBSTANTIATION}\\ \textbf{+BASIS}} &
        \specialcell[c]{The citation both substantiates a claim made in the citing work and serves as the basis for an idea or method\\that is subsequently built upon.} \\
        \vspace{0.01cm}\\
        \specialcell[c]{\textbf{BASIS}} &
        \specialcell[c]{The method, idea, or tool in the citation serves as the basis for the citing work. This must be explicitly stated\\or clearly evident from the abstract that the cited idea is central and explored further.} \\
        \vspace{0.01cm}\\
        \specialcell[c]{\textbf{SUBSTANTIATION}} &
        \specialcell[c]{The citation is used to verify or substantiate theoretical, empirical, or methodological claims made in the citing \\work. When classified as Substantiation, the supported claim should be identifiable.} \\
        \vspace{0.01cm}\\
        \specialcell[c]{\textbf{USE}} &
        \specialcell[c]{The method or tool in the citation is used directly in the citing work, with no or very trivial modifications.}\\
        \vspace{0.01cm}\\
        \specialcell[c]{\textbf{DEFINITION}} &
        \specialcell[c]{The citation is used to define or explain a topic or phrase (but not to validate it).}\\
    
        \vspace{0.01cm}\\
        \specialcell[c]{\textbf{ANALYSIS}} &
        \specialcell[c]{The method or idea in the citation is analyzed through comparisons or critiques that are directly linked to the\\citing work. The cited work may be compared to or criticized to justify an approach taken in the citing work,\\but is not explicitly built upon.} \\

        \vspace{0.01cm}\\
        \specialcell[c]{\textbf{RELATED WORK}} &
        \specialcell[c]{The citation refers to work that is related to the citing work or provides it as an example of a method, idea, or\\phenomenon discussed without further description. It does not fit any of the other categories.} \\
    \bottomrule
     
    \end{tabular}}
    \caption{Citation purpose categories used for annotation.}
    \label{tab:citation-purpose}
\end{table*}

\section{Citation Purpose}
\label{sec:framework}

In this section, we discuss the novel taxonomy developed to classify citation purpose, the corresponding annotation process for obtaining a manually-labeled dataset, and an automatic classifier for citation purpose prediction. 


\subsection{Citation Purpose Taxonomy} 

Measuring the level of interdisciplinary engagement in a publication requires identifying the role that out-of-discipline references play in shaping the paper's conceptual, methodological, and empirical claims. For instance, a citation that contrasts a proposed approach with prior methods in the literature does not necessarily reflect deep engagement. In contrast, citations that reference methods or claims that are foundational to the analyses or arguments advanced in the published work indicate a substantially higher degree of engagement.

In line with this goal, we propose a citation-purpose classification scheme adapted from \citet{abu-jbara-etal-2013-purpose}, who define six categories: \textit{Criticizing}, \textit{Comparison}, \textit{Use}, \textit{Substantiating}, \textit{Basis}, and \textit{Neutral/Other}. Through an iterative, inductive annotation process conducted by two of the authors of the paper, we refine and extend this framework to better capture the roles that citations play in interdisciplinary contexts and to support assessments of engagement depth. The resulting taxonomy is presented in Table~\ref{tab:citation-purpose}. Label determinations are made considering the paragraph surrounding each citation and the claims made in the paper's abstract, which together provide sufficient argumentative context for labeling decisions.

The categories \textit{Basis} and \textit{Substantiation + Basis} point to a deeper engagement with the literature. The latter, in particular, is designed to highlight deep engagement with interdisciplinary work.
Often a claim is validated by an out-of-discipline citation, then used to inform the proposed method. For example, in the sentence 

\begin{quote}
\say{As pronoun usage tends to differ in individuals living with depression \textbf{(Vedula and Parthasarathy, 2017)}, we removed any English pronouns from our stop word set}
\end{quote}

\noindent a citation supporting a claim from the field of psychology directly informs a methodological choice.

Another key feature of our taxonomy is the inclusion of the \textit{Analysis} category, which combines the frequently-used categories \textit{Comparison} and \textit{Criticism}~\cite{abu-jbara-etal-2013-purpose}. In most cases, these categories refer to the same general principle: analyzing the idea in the citation. Whether the ideas are being compared to or criticized does not signal different depths of engagement. The introduction of the \textit{Definition} category helps identify citations that establish core concepts used in the work. Although citing prior work to support a definition may appear surface-level, such citations signal the conceptual importance of the defined term within the work and therefore provide meaningful evidence about the depth of interdisciplinary engagement.

Finally, we specifically design the \textit{Related Work} category to include citations which support claims about the state of the literature (e.g. \say{There exist studies on this subject}, \say{Multiple works have adopted this approach}). This distinction separates such citations from those that support empirical or findings-based claims, which are more appropriately categorized as \textit{Substantiation}. In interdisciplinary work, the latter typically reflects a deeper and more consequential form of engagement. 



\begin{table}[t]
    \begin{center}
    \resizebox{\columnwidth}{!}{
    \begin{tabular}{lccc}
        \toprule
        \textbf{Category} & \textbf{Interdisciplinary} & \textbf{Within Disc.} & \textbf{Total} \\
        \midrule
        Related Work & 163 & 202 & \textbf{365}\\
        Substantiation & 98 & 86 & \textbf{184}\\
        Use & 54 & 115 & \textbf{169}\\
        Analysis & 31 & 77 & \textbf{108}\\
        Basis & 18 & 26 & \textbf{44}\\
        Substantiation + Basis & 26 & 17 & \textbf{43}\\
        Definition & 23 & 14 & \textbf{37}\\
        \bottomrule
    \end{tabular}
    }
    \caption{Number of annotated citations per citation purpose category with full or partial agreement.}
    \label{tab:ann-counts}
    \end{center}
\end{table}

\subsection{Citation Purpose Annotation}
\label{subsec:cpa}
To create our dataset, we performed an annotation study with three of the paper authors and two computer science graduate students. The graduate students received training from the authors and were provided with our Codebook (Appendix \ref{app:codebook}).

We selected 248 articles at random, balanced across publication year and venue. We randomly selected $\sim 1,000$ in-text citations from this set, balancing across the section in which they appear. 624 of the 998 (62.5\%) examples were doubly-annotated, and 374 (37.5\%) were triply-annotated. Agreement statistics were calculated over the entire annotated set (both doubly- and triply-annotated). We had complete agreement on 646 of the citations and partial agreement on 304 (at least partial agreement on 950). We obtained a Krippendorff's $\alpha$ of 0.596. 

The resulting dataset of 950 citations (with at least partial agreement) is summarized in Table \ref{tab:ann-counts}. A pairwise comparison of annotations is included in Figure \ref{fig:disagree}. A comprehensive discussion of common disagreements is included in Appendix \ref{app:cit-purp-agreement}.

\begin{figure}[t]
    \centering
    \includegraphics[width=\linewidth]{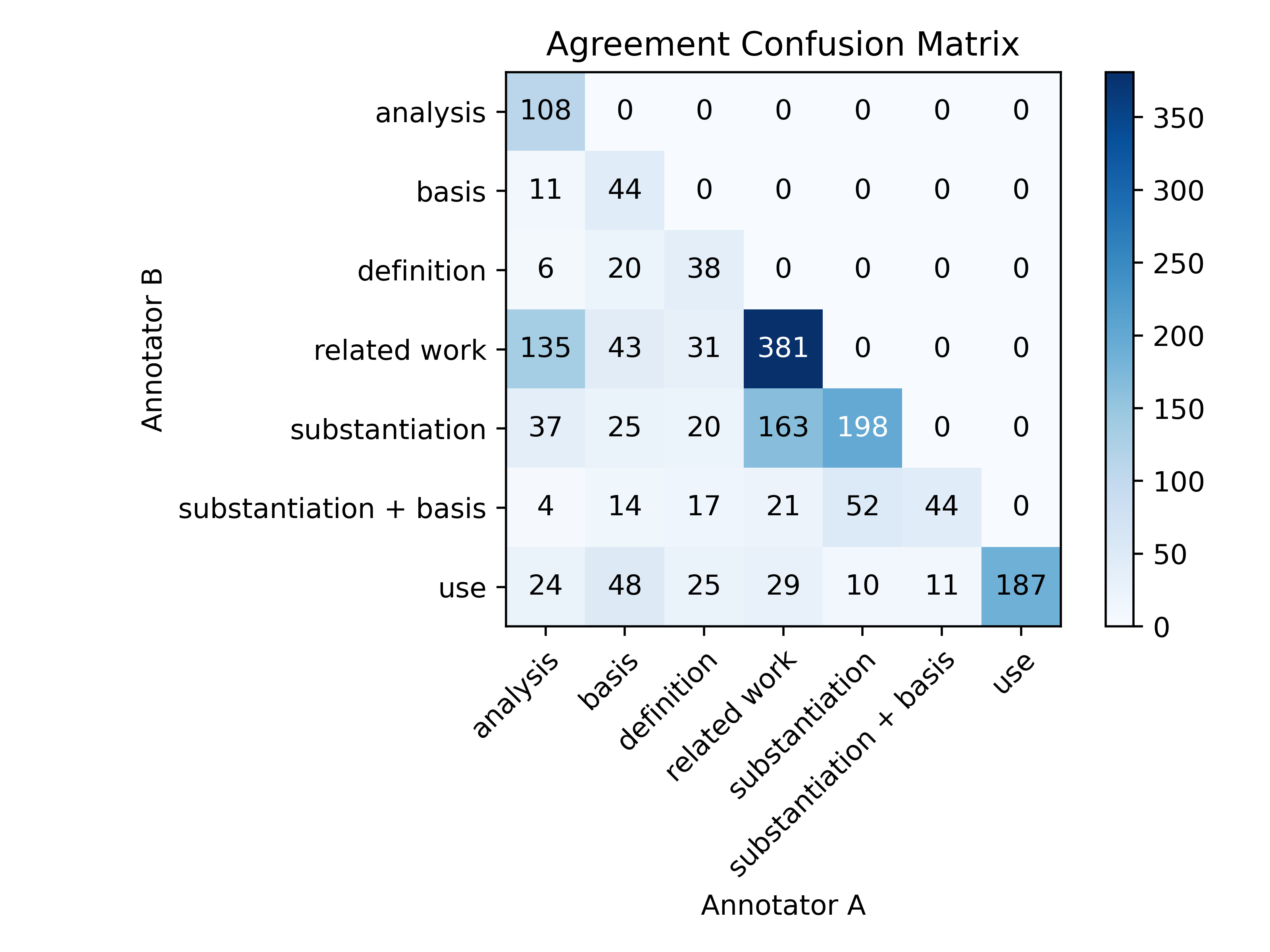}
    \caption{Pairwise comparison of annotations. Each annotation for a sample was considered with each of the other annotations (i.e., there is one pair for doubly-annotated samples and three pairs for triply-annotated samples). Annotator IDs on the x- and y-axes are arbitrary.}
    \label{fig:disagree}
\end{figure}

\subsection{Automatic Citation Purpose Classification}
\label{subsec:cp-classifier}
Here, we present and evaluate SOTA models for automatically classifying citation purpose according to our citation purpose taxonomy. We start by splitting our annotated dataset, at the article level, into a training set consisting of 80\% of the data and a held-out test set of the other 20\% of the data. Though our classifier operates at the in-text citation level, splitting by article guarantees no within-article leakage. We evaluate three models: (1) fine-tuned RoBERTa, (2) prompting \texttt{Llama3-8b} \cite{dubey2024llama3}, and prompting \texttt{Qwen3-8b} \cite{qwen3}. 

We fine-tuned RoBERTa on the training set using the AdamW optimizer, weighted cross-entropy loss, and a learning rate of $4e-5$. For early stopping, we use the macro F1 on the validation set (10\% of the training set). Temperature was set to 1 for \texttt{Llama3-8b} and \texttt{Qwen3-8b}; prompts are provided in Appendix \ref{app:prompts}. We report precision, recall, weighted f1, and macro f1 performance on the held-out test set in Table \ref{tab:model-results}. We include detailed classification reports for each model in Appendix \ref{app:cit-purp-addl-results}.

\paragraph{Results}
We find that \texttt{Qwen3} out-performs all other models, achieving the greatest precision, recall, weighted f1-score, and macro f1-score. We use our classifier to automatically obtain Citation Purpose classifications across the entire unlabeled dataset used for the case study in Section \ref{sec:case-study}.

\begin{table}[]
    \centering
    \resizebox{\columnwidth}{!}{
    \begin{tabular}{l|rrrr}
        \toprule
        Model & Precision & Recall & Weighted F1 & Macro F1 \\
        \midrule
        Random & 0.18 & 0.18 & 0.22 & 0.16\\
        \midrule
        \texttt{Llama3} & 0.30 & 0.25 & 0.19 & 0.15\\
        \texttt{Qwen3} & \textbf{0.58} & \textbf{0.58} & \textbf{0.60} & \textbf{0.56} \\
        \texttt{RoBERTa} & 0.48 & 0.50 & 0.60 & 0.49\\
        \bottomrule
    \end{tabular}
    }
    \caption{Automatic citation purpose classification results.}
    \label{tab:model-results}
\end{table}

\section{Citation Engagement Predictor}
\label{sec:engagment-score}
We posit that the depth of engagement reflected by a citation is captured by two components: the  purpose and the section of its in-text references.
In this section we present a Citation Engagement Predictor (CEP) which seeks to combine these components for a singular rating of a publication’s engagement with a given citation. We define a 5-point scale for this rating and present a method for predicting it automatically.

\subsection{Citation Engagement Annotation Study}
\label{subsec:ceas}
We run an annotation study to ground our CEP in expert judgments. We randomly sample 100 article-citation pairs, distributed as evenly as possible by frequency of occurrence of the citation in the article and by article year and venue, to be representative of the full dataset. Two computer science researchers with expertise in NLP+CSS work analyzed how each cited work was used throughout the article, considering all of its in-text references, and assigned it an engagement score on a five-point Likert scale. A score of one indicates that the annotator \say{strongly disagrees} with the statement \say{The article engages deeply with the cited work}, and five indicates \say{strongly agrees}. On a first pass, we achieved high inter-annotator agreement (ordinal Krippendorff's $\alpha$ of 0.84). The thirty resulting disagreements tended to occur with only a one-point difference. The annotators deliberated on disagreements to arrive at a final set of annotations. For each of these 100 cited works, we also had the same graduate students hand-annotate the purpose of each citation occurrence, and obtained section labels from our data collection pipeline (Section \ref{sec:data}).

\subsection{Modeling Engagement}
For each cited work, we construct features representing how often it is referenced with each purpose and in each section. We fit an ordinal logistic regression model on our 100 annotated cited works to learn the boundaries between engagement levels and the contribution of each feature to engagement. We use 7 purpose category features and 6 section categories for a total of 13 features. 

Because every in-text reference is assigned exactly one purpose label and appears in exactly one section, the 7 purpose counts and the 6 section counts for each cited work both sum to its total number of in-text references. This creates a linear dependency between the purpose and section features, which prevents the model from uniquely estimating their coefficients. To address this, we drop one category from each dimension as the baseline. We choose ``Related Work'' in both cases, since it intuitively signals the least engagement and keeps the remaining coefficients positive and easier to interpret. This leaves us with 11 features. We set the value of each feature to  $log(1 + n_x)$, where $n_x$ is the number of in-text references with that purpose or in that section. 

We perform a 90-10 train-test split on our set of 100 article-citation pairs, at the article level, to ensure that the results were not skewed by within-article leakage. A 10-fold cross validation on our model showed low variance and standard deviation across folds, providing confidence that the model is reliable and generalizable. 

\begin{figure}
    \centering
    \includegraphics[width=\linewidth]{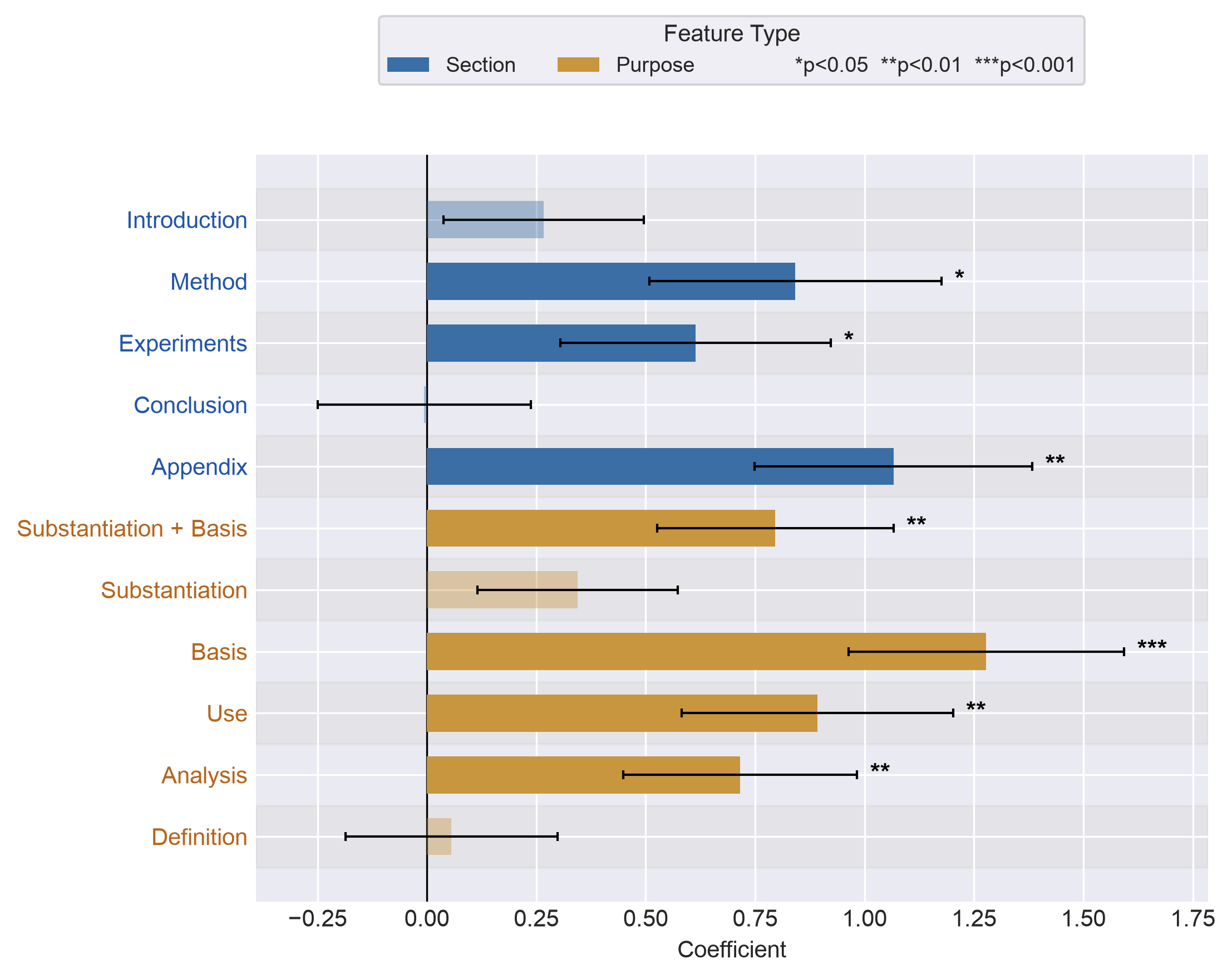}
    \caption{Regression coefficients per feature.}
    \label{fig:regression-coefficients}
\end{figure}


\subsection{Results}

We achieve a Mean Absolute Error (MAE) of $0.476^{\pm 0.110}$, a Quadratic Weighted Kappa (QWK) of $0.813^{\pm 0.064}$, and an accuracy of $0.563^{\pm 0.129}$, averaged across the 10 folds. We report coefficients in Figure \ref{fig:regression-coefficients}, which tell us the increase in log-odds of being rated at a higher engagement level as the citation counts per section or purpose increase. We see that having the purpose \textit{Basis} and being in the \textit{Appendix} section are especially indicative of a citation being engaged with highly, with low p-values. Notably, the p-values of the coefficients for the \textit{Definition} purpose category and the \textit{Conclusion} section are particularly high, suggesting that more data is required to properly draw conclusions about the effects of these features on engagement levels. In general, we find that the purpose of a citation, i.e., why a citation is used, offers more insight into its level of engagement in an article than where it occurs. This finding underscores the value of our extension to \citet{leto-etal-2024-first}. While section alone carries some signal, the purpose of a citation (as captured by our taxonomy) is the more informative component, and incorporating it yields a substantially richer measure of engagement. 

\begin{table}[]
    \centering
    \begin{tabular}{l|rr}
        \toprule
        Features & MAE & QWK \\
        \midrule
        All & 0.476 ± 0.110 & 0.813 ± 0.064 \\
        Section only & 0.585 ± 0.070 & 0.749 ± 0.108 \\
        Purpose only & 0.556 ± 0.116 & 0.716 ± 0.112 \\
        \bottomrule
    \end{tabular}
    \caption{Model performance with only a subset of features.}
    \label{tab:ablation-feature-subset}
\end{table}

\begin{table}[]
    \centering
    \begin{tabular}{l|rr}
        \toprule
        Features Dropped & LR & p-value \\
        \midrule
        Purpose & 30.3 & 0.00003 \\
        Section & 16.4 & 0.005 \\
        \bottomrule
    \end{tabular}
    \caption{Log-likelihood ratios with features dropped in comparison to the full feature set.}
    \label{tab:ablation-log-likelihood-ratios}
\end{table}

We conduct ablation studies to further confirm these findings and validate our choice to use both sets of features. Table \ref{tab:ablation-feature-subset} shows the MAE and QWK when only a subset of features is used. The lower performance in both cases indicates that both the section and purpose feature sets are necessary and informative. Table \ref{tab:ablation-log-likelihood-ratios} shows the log-likelihood ratios with features dropped. The high ratios with respect to the full set indicate that the improvement from the addition of features is not just chance and arises from a substantial improvement in predictive power. It is also clear from the latter experiment that purpose is the stronger predictor, as removing purpose features results in a substantially larger drop in log-likelihood.

\section{NLP + CSS Engagement Analysis}
\label{sec:case-study}
In this section, we analyze our dataset to demonstrate the utility of our proposed Citation Purpose framework and Citation Engagement Predictor. We leverage the automatic Citation Purpose classifier (Section \ref{sec:framework}) to obtain citation purpose classifications for all in-text citations in our dataset. Then, we use our Citation Engagement Predictor to obtain the level of engagement (Section \ref{sec:engagment-score}) for each citation-article pair in our dataset, considering all occurrences of the citation within the article, and their purposes and sections. We drop pairs which do not have purpose or section information for all occurrences. While individual purpose predictions and engagement estimates carry some error, we expect aggregate trends across our dataset of 34956 article-citation pairs to be informative. We present these analyses below.

\begin{table}[t]
    \begin{center}
    \resizebox{0.65\columnwidth}{!}{
    \begin{tabular}{lccc}
        \toprule
        \textbf{Level} & \textbf{Mean} & \textbf{Median} & \textbf{Maximum} \\
        \midrule
        1 & 11.05 & 9 & 78 \\
        2 & 4.6 & 4 & 44 \\
        3 & 1.19 & 1 & 16 \\
        4 & 0.22 & 0 & 6 \\
        5 & 0.18 & 0 & 5 \\
        \bottomrule
    \end{tabular}
    }
    \caption{Number of citations per engagement level per article.}
    \label{tab:num-citations-per-engagement-level}
    \end{center}
\end{table}

\subsection{Quantitative Analysis}

\noindent\textbf{Articles have more low-engagement citations.} Out of all article-citation pairs, we find that 64.1\% (22398) have Level 1 (low) engagement, 26.7\% (9328) and 6.9\% (2416) have Levels 2 and 3 (medium) engagement, and only 1.3\% (443) and 1.1\% (371) have Levels 4 and 5 (high) engagement. We present statistics on the number of citations per engagement level per article in Table \ref{tab:num-citations-per-engagement-level}, finding that papers tend to engage with only a few citations on a deeper level.

\begin{figure}
    \centering
    \includegraphics[width=\linewidth]{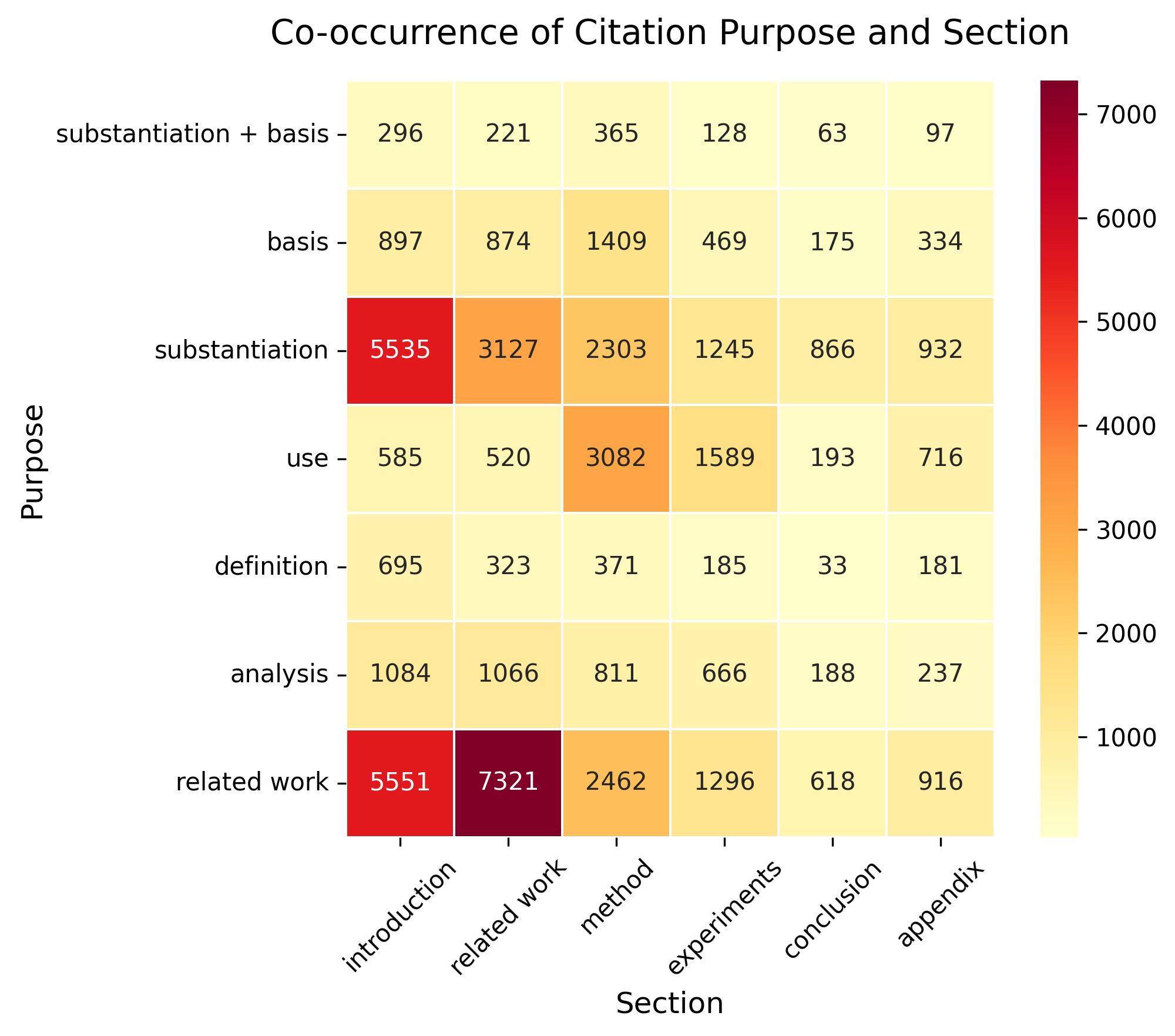}
    \caption{Co-occurrence of citation purpose and section in full dataset.}
    \label{fig:purpose-section-cooccurrence}
\end{figure}

\begin{figure}
    \centering
    \includegraphics[width=\linewidth]{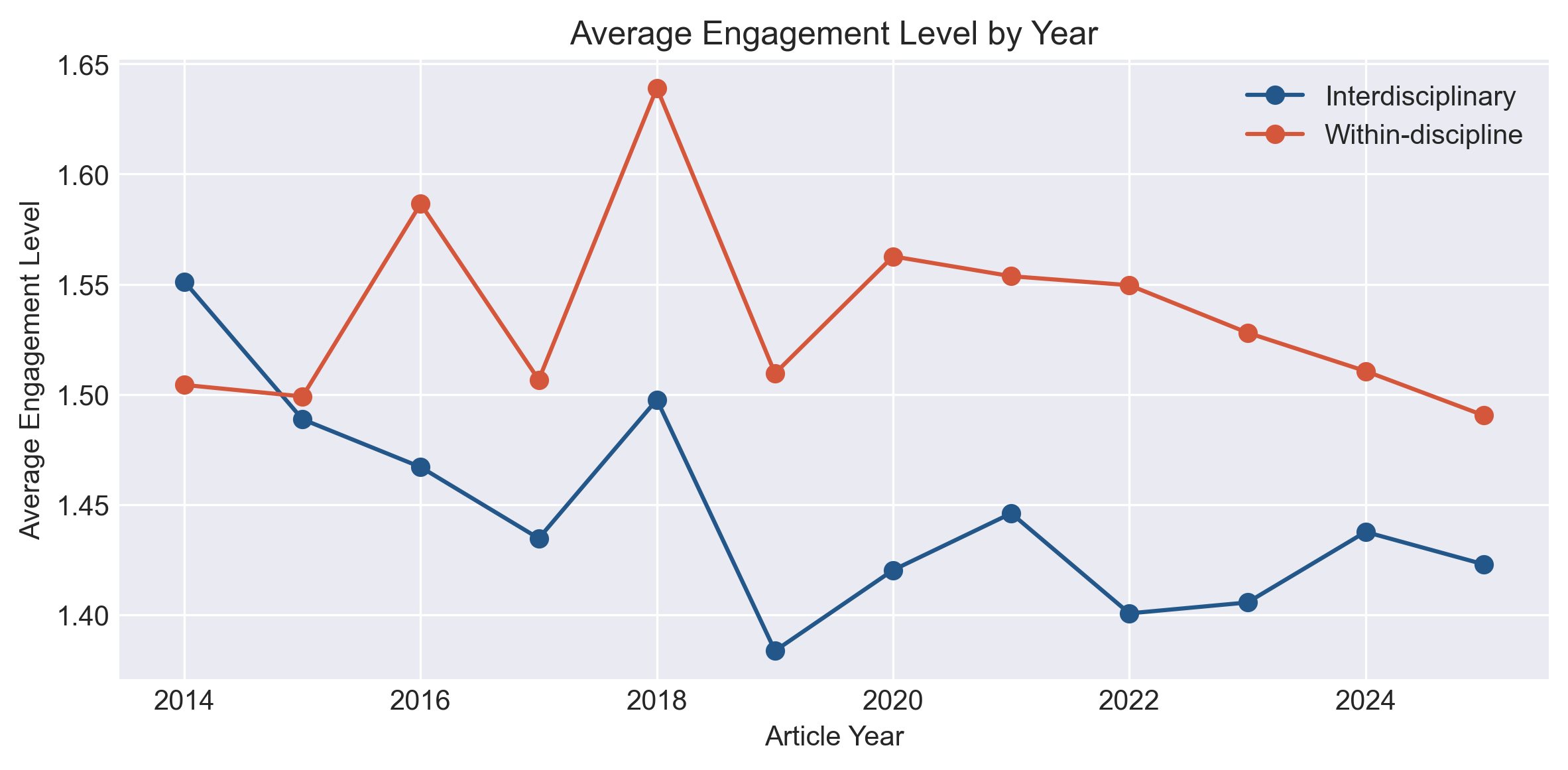}
    \caption{Average engagement level by article year.}
    \label{fig:engagement-level-year}
\end{figure}

\paragraph{Some sections and citation purposes intuitively co-occur.} To predict the engagement level, we use the purposes of the citation in the article and the sections of occurrence; the co-occurrence between these features on our complete dataset, at the in-text citation level, is in Figure \ref{fig:purpose-section-cooccurrence}. We find that many citations used for the purpose of \textit{Related Work} often appear in the \textit{Introduction} and \textit{Related Work} sections. \textit{Substantiation} also often occurs in the \textit{Introduction}, while \textit{Use}-purpose citations are common in the \textit{Method} and \textit{Experiments} sections. However, we see enough variation in these patterns to require both components to predict engagement. We verify that these trends hold in the subset of citations hand-annotated with citation purpose (see Appendix Figure \ref{fig:ann-sec-purp-cooccur}).

\paragraph{Articles show deeper engagement with within-discipline citations.}
In Figure \ref{fig:engagement-level-year}, we show the change in average engagement level by year. We find that authors tend to engage more deeply with within-discipline citations than interdisciplinary citations. However, this trend does not hold for the year 2014, although the difference is minimal.

\begin{figure}
    \centering
    \includegraphics[width=\linewidth]{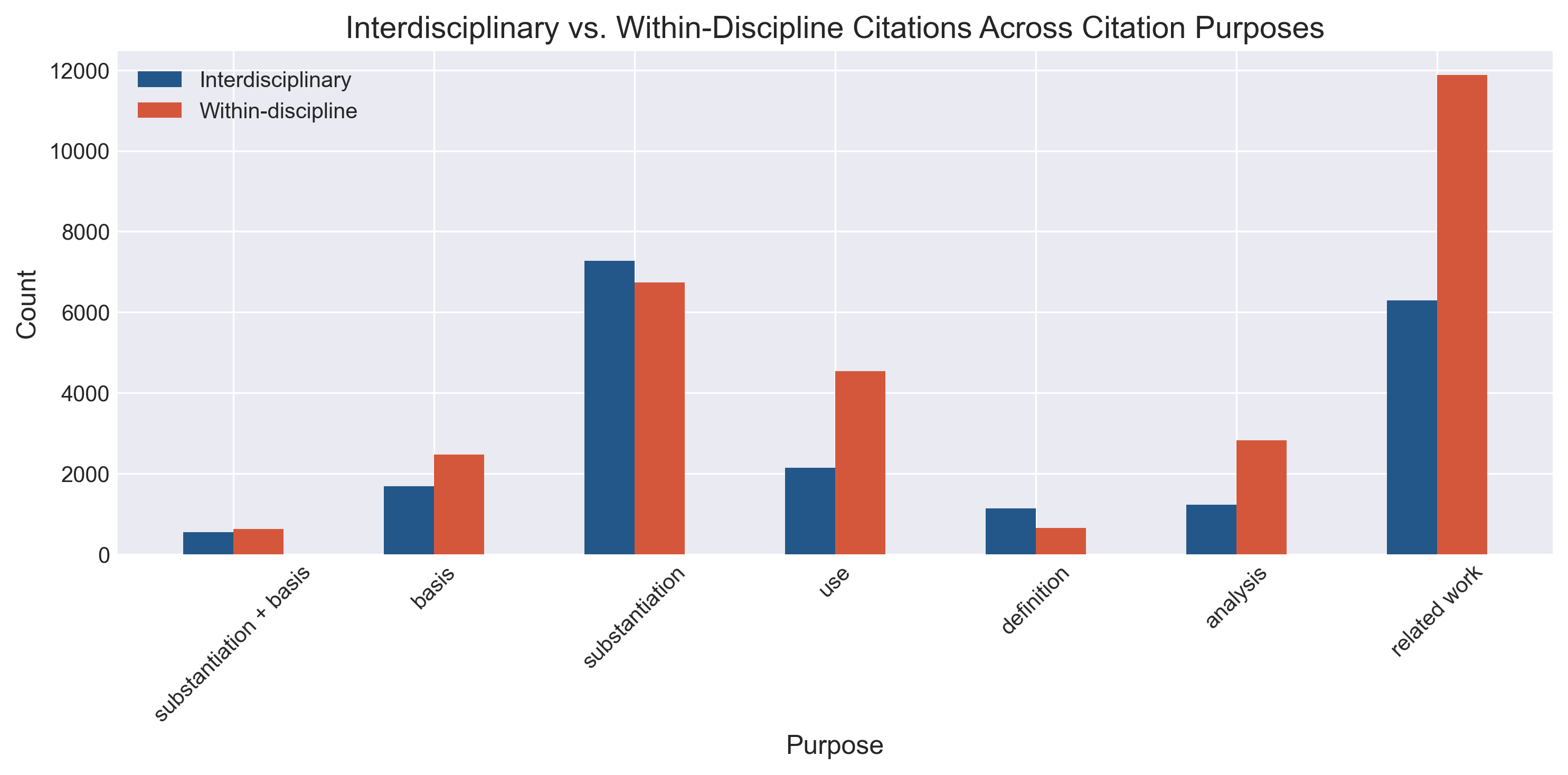}
    \caption{Interdisciplinary vs. Within-Discipline citations by citation purpose.}
    \label{fig:interdisciplinary-citation-purpose}
\end{figure}

\paragraph{Certain purposes are more associated with interdisciplinary citations.} An analysis of the purpose counts of within-discipline vs. out-of-discipline citations is in Figure \ref{fig:interdisciplinary-citation-purpose}. This plot shows that the categories \textit{Substantiation} and \textit{Definition} in particular are more frequent for interdisciplinary references. This is also reflected in the subset of citations hand-annotated with citation purpose (see Appendix Figure \ref{fig:ann-in-vs-out-eng}).

\subsection{Qualitative Analysis}


\begin{table*}[t]
    \begin{center}
    \renewcommand{\arraystretch}{1.3}
    \begin{tabularx}{\linewidth}{Xcc}
        \toprule
        \multirow{2}{*}{\textbf{Paper}} & \textbf{High-Eng.} & \textbf{Majority}\\
        ~ & \textbf{Citations} & \textbf{Interdisc.?} \\
        \midrule
        \say{When do Word Embeddings Accurately Reflect Surveys on our Beliefs About People?} \cite{joseph-morgan-2020-word} & 4 & Yes\\
        
        \say{Modeling and Prediction of Online Product Review Helpfulness: A Survey} \cite{ocampo-diaz-ng-2018-modeling} & 8 & No\\
        
        \say{Ideology Prediction from Scarce and Biased Supervision: Learn to Disregard the `What' and Focus on the `How'!} \cite{chen-etal-2023-ideology} & 2 & Yes\\
        
        \say{Personality Matters: User Traits Predict LLM Preferences in Multi-Turn Collaborative Tasks} \cite{yunusov-etal-2025-personality} & 0 & N/A\\
        \say{Deciphering Oracle Bone Language with Diffusion Models} \cite{guan-etal-2024-deciphering} & 0 & N/A\\
        \bottomrule
    \end{tabularx}
    \caption{Articles used for qualitative analysis.}
    \label{tab:qual-analysis-articles}
    \end{center}
\end{table*}

We conduct a qualitative analysis of a set of articles to understand how various engagement levels with cited works manifest in a complete paper (see Table \ref{tab:qual-analysis-articles}). We select two articles that have 4+ high-engagement citations (levels 4 and 5) --- one where the majority of these are interdisciplinary, and one where they are not. We compare these to articles which have only 1-2 high-engagement citations, as well as articles which have no high-engagement citations.

The article \say{When Do Word Embeddings Accurately Reflect Surveys on our Beliefs About People?} \cite{joseph-morgan-2020-word}, has 4 high-engagement citations. They 
engage with multiple citations ('Kozlowski et al., 2019', 'Swinger et al., 2019'), to motivate their design of embedding concepts to recover beliefs, making their adoption of concepts from another discipline --- cognitive science --- principled and well-motivated. Their experiments on specific belief taxonomies proposed by other high-engagement citations ('Smith-Lovin and Robinson, 2015'), add further rigor to their contributions. 

The article \say{Personality Matters: User Traits Predict LLM Preferences in Multi-Turn Collaborative Tasks} \cite{yunusov-etal-2025-personality} similarly engages with cognitive traits, but our framework assigns it lower engagement scores across its citations, interdisciplinary and otherwise. The authors adopt a taxonomy from the cognitive science literature, but the citation context does not extensively motivate why this taxonomy is the right tool for their analysis. In contrast, \citet{joseph-morgan-2020-word} examine the different methods of constructing dimensions, across sociocultural themes, before selecting a set; this provides additional grounding for its methodological choices. 

Similarly, \say{Ideology Prediction from Scarce and Biased Supervision: Learn to Disregard the `What' and Focus on the `How'!} \cite{chen-etal-2023-ideology} has a small number of high-engagement citations and tends to anchor one or two central aspects of its work in them. While the article offers a thoughtful justification for this foundation, the depth of citation engagement is concentrated in those few aspects rather than distributed across the analysis, as is more common in articles with multiple high-engagement citations.

\say{Modeling and Prediction of Online Product Review Helpfulness: A Survey} \cite{ocampo-diaz-ng-2018-modeling} provides an example of an article that has many high-engagement, within-discipline citations. In this article, we observe that the high- and medium-engagement citations (`Tang et al., 2013', `Yang et al., 2015') describe datasets that are central to its work, and are engaged with in multiple ways. 
In contrast, the article \say{Deciphering Oracle Bone Language with Diffusion Models} \cite{guan-etal-2024-deciphering}, despite being closely tied to a historical writing system, does not deeply engage with any of the related work from that field, choosing instead to focus on the methodological improvement within the discipline of Computer Science. 

In general, we find that interdisciplinary citations tend to be engaged with more at the idea and claim level, whereas within-discipline citations are engaged with methodologically. We see a similar pattern when comparing articles with many high-engagement citations to those with only one or two. We also find that high-engagement citations tend to span multiple sections and purposes, reflecting how deeply integrated work is woven throughout a paper. These patterns are not captured by conventional measures such as citation entropy or diversity indices, which treat all citations as equivalent. By capturing \textit{how} authors engage with the work they cite, rather than only \textit{what} they cite, our framework provides a complementary view to existing measures of interdisciplinarity.

\section{Conclusion}

Motivated by the need to evaluate the engagement of interdisciplinary publications, we present a novel annotation taxonomy designed to capture citation purpose. The taxonomy includes categories specifically designed to capture interdisciplinary citation purpose. We present an annotation study of NLP+CSS work and a resulting dataset of 950 agreed-upon annotated citations, along with an automatic classifier for predicting citation purpose. We operationalize our Citation Purpose taxonomy to present a Citation Engagement Predictor for estimating a publication's depth of engagement with a given citation given signal from the section and purpose of corresponding in-text references. Finally, we demonstrate the usefulness of our resulting framework by presenting an analysis of a set of NLP+CSS publications, finding that only a minority engage very deeply with the works they cite. Because we develop and analyze this framework in the context of NLP work, it serves as a first step for the development of a broader framework to study interdisciplinarity. We leave the validation of this framework in additional fields to future work.


\section*{Limitations}
We offer an overview of the failure points of our framework in Appendix \ref{app:error-analysis}. However, we would like to emphasize a few limitations here: (1) An automated framework was used to construct the dataset, potentially resulting in parsing errors and inaccuracies when extracting elements from scientific publications. (2) The presented analysis spans a limited number of articles and citations and includes only NLP+CSS work. Further analysis is required to understand whether the takeaways generalize to a larger dataset or additional fields of study. (3) An automatic classifier was used to predict the Citation Purpose labels used for the case study in Section \ref{sec:case-study}. As a result, the analysis carries some uncertainty.

\section*{Acknowledgments}

This work utilized the Blanca high performance computing resources at the University of Colorado Boulder. Blanca is jointly funded by computing users and the University of Colorado Boulder.

\bibliography{latex/custom}

\appendix

\section{Additional Data Collection Details}
\label{app:data-collect-details}

Here we include additional details for our interdisciplinary article collection and citation mapping processes. 

\begin{figure}
    \centering
    \includegraphics[width=\linewidth]{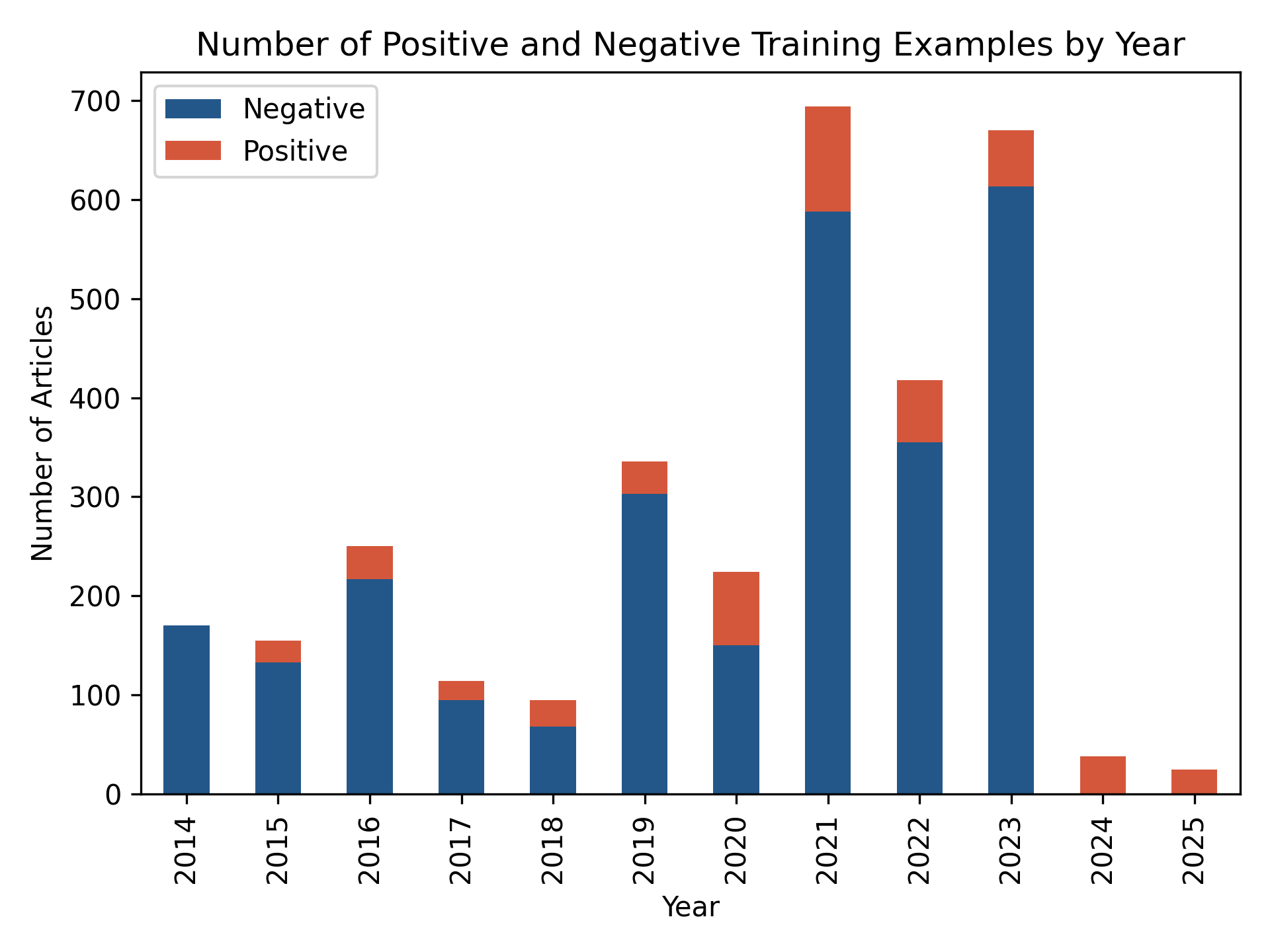}
    \caption{Labeled dataset for training track classifier}
    \label{fig:track_labels_year}
\end{figure}

\subsection{Automatic Track Classifier Details}
\label{app:track-classifier}
To build our binary track classifier, we fine-tune RoBERTa base \cite{gururangan-etal-2020-dont} on the labeled abstracts (Figure \ref{fig:track_labels_year}) using HuggingFace's Transformers Trainer \cite{wolf2020huggingfacestransformersstateoftheartnatural}, employing a weighted cross-entropy loss to address class imbalance. The model is trained for 50 epochs with a learning rate of $2e-5$, using an 80/10/10 train/validation/test split. The checkpoint achieving the highest macro F1 score on the validation set is selected, yielding a macro F1 of 0.870, precision of 0.863 and recall of 0.876 on the test set. 

\subsection{Mapping In-text References to Sections}
\label{app:sec-mapping}
To arrive at the canonical section, the section titles initially parsed by Grobid are assigned to one of \say{Introduction,} \say{Related Work,} \say{Method,} \say{Experiments,} \say{Conclusion,} or \say{Appendix} using string-matching rules. The first section is always mapped to \say{Introduction.} Then, additional sections are mapped to titles if they contain the keywords shown in Table \ref{tab:section-keywords}. Sections that cannot be mapped are not used for analysis.

\begin{table*}[t]
\centering
    \begin{tabular}{lp{0.75\textwidth}}
        \toprule
        \textbf{Section} & \textbf{Keywords} \\
        \midrule
        Related Work & \say{related work,} \say{background,} \say{related research} \\
        \addlinespace
        Method & \say{method,} \say{approach,} \say{notation,} \say{technique,} \say{algorithm,} \say{architecture,} \say{design,} \say{solution,} \say{model,} \say{corpus,} \say{data,} \say{framework} \\
        \addlinespace
        Experiments & \say{experiment,} \say{evaluation,} \say{test,} \say{analysis,} \say{vs.,} \say{compare,} \say{accuracy,} \say{scores,} \say{state-of-the-art,} \say{baseline,} \say{results,} \say{performance} \\
        \addlinespace
        Conclusion & \say{conclusion,} \say{future work,} \say{discussion,} \say{limitation,} \say{ethical,} \say{ethics} \\
        \addlinespace
        Appendix & \say{appendix,} \say{appendices} \\
        \bottomrule
    \end{tabular}
\caption{Keywords used for matching section titles to canonical sections.}
\label{tab:section-keywords}
\end{table*}

\section{Citation Purpose Classifier Details}
\label{app:cit-purp-addl-results}
In this section we include full classification reports for \texttt{Llama3} in Table \ref{tab:llama-classification-report}, \texttt{Qwen3} in Table \ref{tab:qwen-classification_report}, and \texttt{RoBERTa} in Table \ref{tab:rob-classification_report}.




\begin{table}[t]
    \begin{center}
    \resizebox{\columnwidth}{!}{
    \begin{tabular}{lrrrr}
        \toprule
        \textbf{Class} & \textbf{Precision} & \textbf{Recall} & \textbf{F1-Score} & \textbf{Support} \\
        \midrule
        substantiation + basis & 0.00 & 0.00 & 0.00 & 10 \\
        substantiation         & 0.50 & 0.39 & 0.44 & 44 \\
        basis                  & 0.07 & 0.38 & 0.12 & 8  \\
        definition             & 0.05 & 0.67 & 0.10 & 6  \\
        analysis               & 0.44 & 0.29 & 0.35 & 24 \\
        use                    & 0.00 & 0.00 & 0.00 & 19 \\
        related work           & 1.00 & 0.02 & 0.03 & 57 \\
        \midrule
        accuracy               &      &      & 0.19 & 168 \\
        macro avg              & 0.30 & 0.25 & 0.15 & 168 \\
        weighted avg           & 0.54 & 0.19 & 0.19 & 168 \\
        \bottomrule
    \end{tabular}
    }
    \caption{\texttt{Llama3} classification report showing precision, recall, and F1-score per citation purpose category.}
    \label{tab:llama-classification-report}
    \end{center}
\end{table}

\begin{table}[h]
\centering
    \resizebox{\columnwidth}{!}{
        \begin{tabular}{lrrrr}
            \toprule
            \textbf{Class} & \textbf{Precision} & \textbf{Recall} & \textbf{F1-Score} & \textbf{Support} \\
            \midrule
            substantiation + basis & 0.20 & 0.20 & 0.20 & 10 \\
            substantiation         & 0.67 & 0.50 & 0.57 & 44 \\
            basis                  & 0.38 & 0.75 & 0.50 & 8  \\
            definition             & 0.80 & 0.67 & 0.73 & 6  \\
            analysis               & 0.57 & 0.33 & 0.42 & 24 \\
            use                    & 0.79 & 0.79 & 0.79 & 19 \\
            related work           & 0.65 & 0.81 & 0.72 & 57 \\
            \midrule
            accuracy               &      &      & 0.61 & 168 \\
            macro avg              & 0.58 & 0.58 & 0.56 & 168 \\
            weighted avg           & 0.62 & 0.61 & 0.60 & 168 \\
            \bottomrule
        \end{tabular}
    }
\caption{\texttt{Qwen3} classification report showing precision, recall, and F1-score per citation purpose category.}
\label{tab:qwen-classification_report}
\end{table}

\begin{table}[h]
\centering
    \resizebox{\columnwidth}{!}{
        \begin{tabular}{lrrrr}
            \toprule
            \textbf{Class} & \textbf{Precision} & \textbf{Recall} & \textbf{F1-Score} & \textbf{Support} \\
            \midrule
            substantiation + basis & 0.00 & 0.00 & 0.00 & 10 \\
            substantiation         & 0.67 & 0.70 & 0.69 & 44 \\
            basis                  & 0.22 & 0.25 & 0.24 & 8  \\
            definition             & 0.60 & 0.50 & 0.55 & 6  \\
            analysis               & 0.40 & 0.33 & 0.36 & 24 \\
            use                    & 0.86 & 1.00 & 0.93 & 19 \\
            related work           & 0.63 & 0.72 & 0.67 & 57 \\
            \midrule
            accuracy               &      &      & 0.62 & 168 \\
            macro avg              & 0.48 & 0.50 & 0.49 & 168 \\
            weighted avg           & 0.58 & 0.62 & 0.60 & 168 \\
            \bottomrule
        \end{tabular}
    }
    \caption{\texttt{RoBERTa} classification report showing precision, recall, and F1-score per citation purpose category.}
    \label{tab:rob-classification_report}
\end{table}

\section{Additional Citation Purpose Annotation Details}
\label{app:citation_purpose_distribution_details}
\begin{figure}
    \centering
    \includegraphics[width=\linewidth]{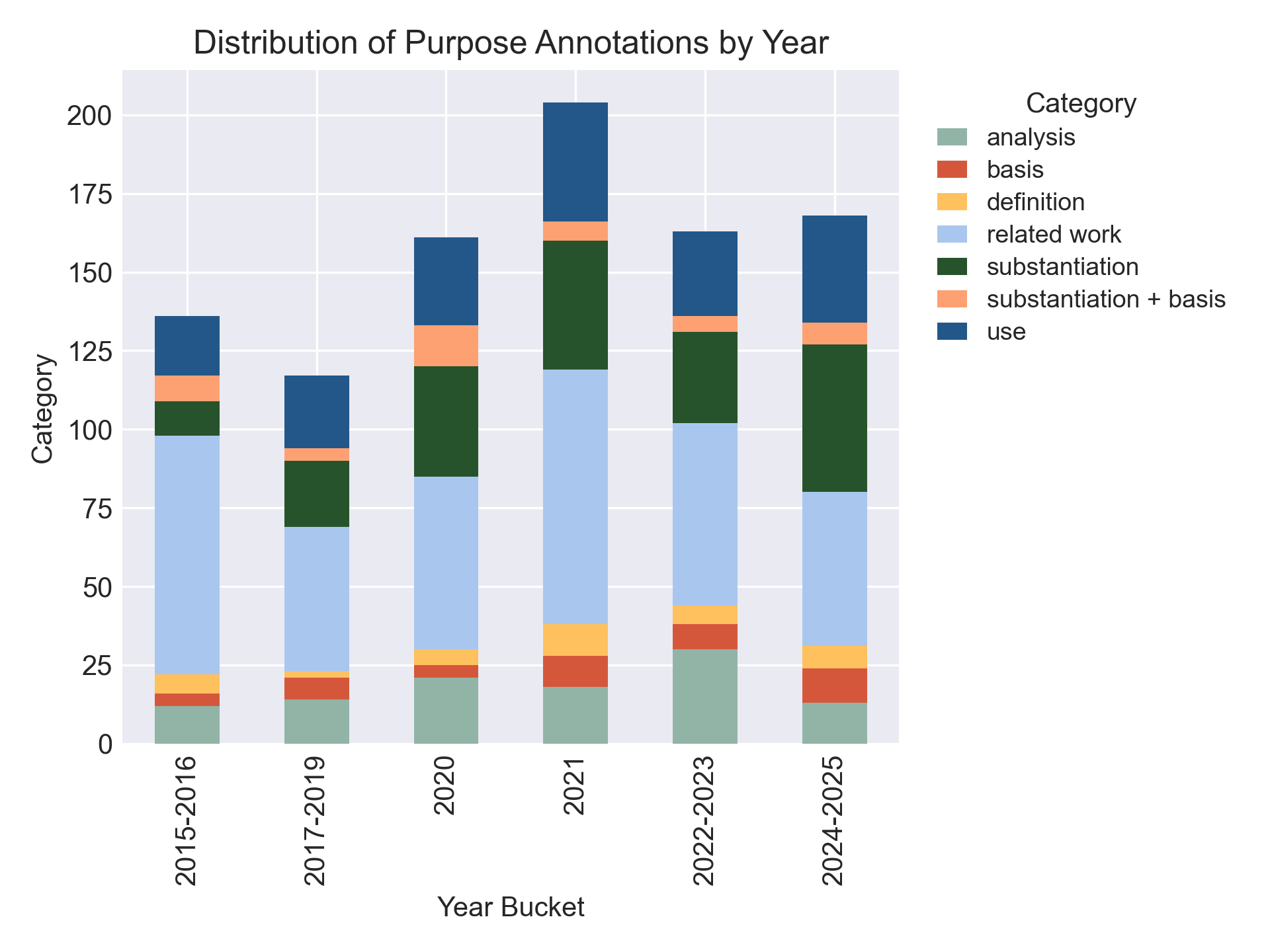}
    \caption{Distribution of agreed-upon citation purpose annotations by article year.}
    \label{fig:purp-dist-year}
\end{figure}

\begin{figure}
    \centering
    \includegraphics[width=\linewidth]{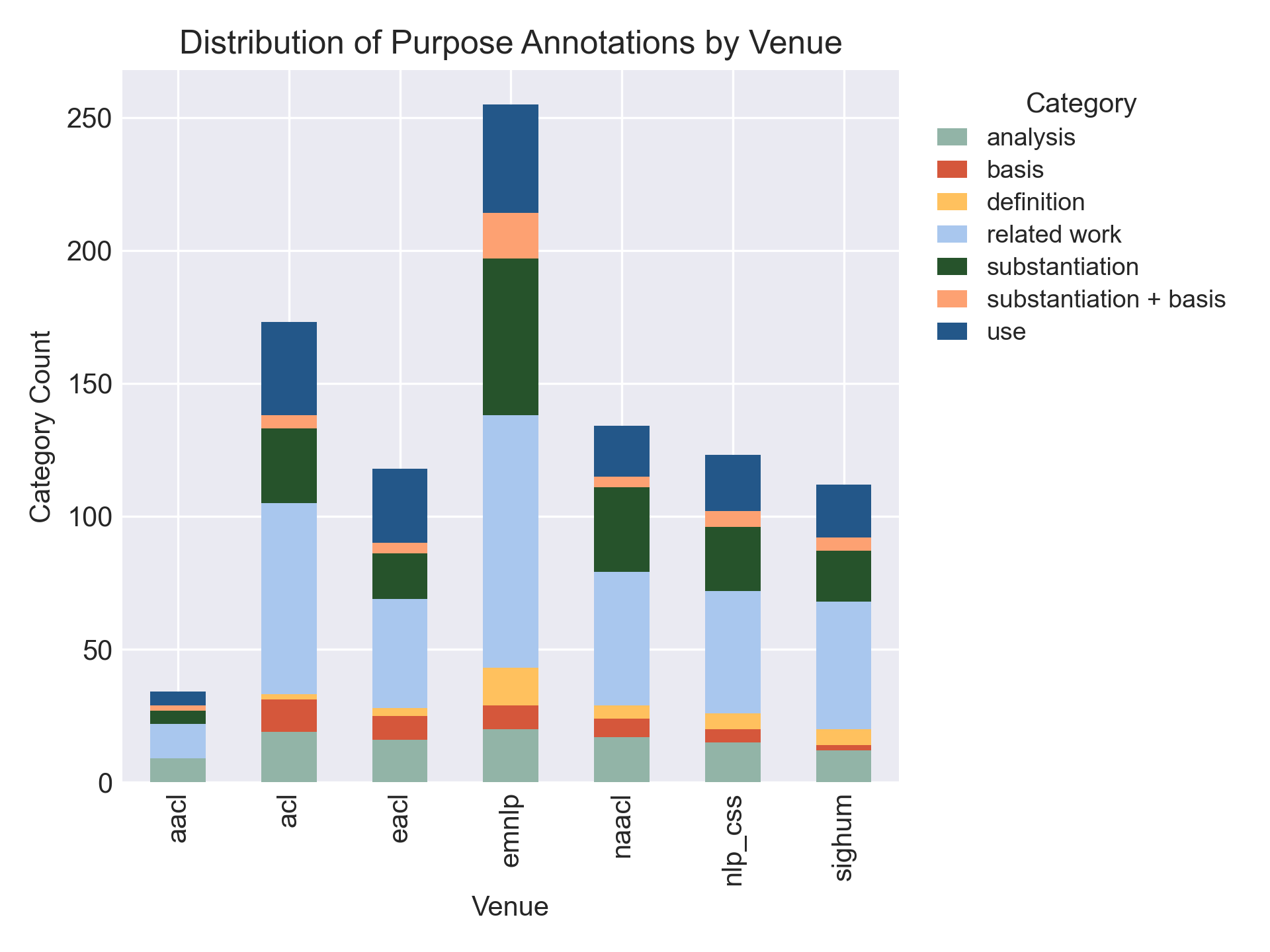}
    \caption{Distribution of agreed-upon citation purpose annotations by article venue.}
    \label{fig:purp-dist-venue}
\end{figure}

Here, we present statistics on the distribution of citation purpose within the manually-annotated dataset across the citing work's year bucket (Figure \ref{fig:purp-dist-year}) and venue (Figure \ref{fig:purp-dist-venue}).

\section{Common Disagreements in Citation Purpose Annotations}
\label{app:cit-purp-agreement}

Here we discuss common disagreements in Citation Purpose annotation. Figure \ref{fig:disagree} shows all disagreements from our annotations study. Below we outline all common disagreements. 

\paragraph{Related Work \& Analysis}
There is some confusion between \textit{Related Work} and \textit{Analysis}. Commonly, this is because a comparison to the citing work, which would constitute \textit{Analysis}, is subtle. Consider the below example of one such disagreement, taken from \citet{hardalov-etal-2020-exams}, where the in-text reference in question is shown in bold:  

\begin{quote}
    \textit{Knowing that the context retrieved from the noisy Wikipedia corpus is relevant for answering Exodus questions, suggests that we need a better multilingual science corpus, similar to Clark et al. (2018); \textbf{Pan et al. (2019)}; Bhakthavatsalam et al. (2020). We further need better multilingual knowledge selection and ranking (Banerjee et al., 2019).} 
\end{quote}

While the authors are indeed making a positive comparison to a pre-existing corpus, it is easily overlooked. We note that positive comparisons are somewhat more difficult to detect than negative ones.

\paragraph{Related Work \& Substantiation}
There is also some confusion between \textit{Related Work} and \textit{Substantiation}. This is generally due to less obvious claims, making them difficult to identify (as suggested in Part III of the codebook in Appendix \ref{app:codebook-tb}). In the below excerpt from \citet{masud-etal-2024-hate}, the reference is used to substantiate one part of a larger claim:

\begin{quote}
    \textit{Disparity in access to additional context (Ljubešić et al., 2022;İhtiyar et al., 2023) and annotation guidelines \textbf{(Ross et al., 2016)}, in some instances, reduce the bias and, in some cases, confirm the annotator's biases.}
\end{quote}

We argue that this could be easily mistaken for an example, constituting a \textit{Related Work} classification.

\paragraph{Substantiation \& Substantiation + Basis}
Confusion between \textit{Substantiation} and \textit{Substantiation + Basis} is also relatively common. This is relatively intuitive, as it's relatively simple to identify the claim constituting a \textit{Substantiation} classification. However, language suggesting that this claim is explicitly being \textit{built on} is subtler. Take the below excerpt from \citet{desai-etal-2020-detecting} as an example.

\begin{quote}
    \textit{Domain adaptation techniques address these challenges by efficiently using large amounts of unlabeled target domain data, consequently outperforming the aforementioned supervised techniques (\textbf{Alam et al., 2018};Li et al., 2017). Our work contributes to disaster-centric emotion detection in three ways by: (1) introducing a dataset large enough to train supervised classifiers; (2) exploring various forms of pre-training to instill strong inductive biases...}
\end{quote}

The claim that \say{domain adaptation techniques outperform prior supervised techniques} is relatively straightforward to identify. However, the citing work also explicitly aims to build on the technique(s) proposed in \textit{Alam et al., 2018} to make them more effective, constituting a \textit{Substatiation + Basis} classification.

\section{Replicating Engagement Analysis Results on Annotated Data}
\label{app:rep-results}

Because the the analysis in Section \ref{sec:case-study} is conducted over automatic citation purpose \textit{predictions}, we replicate results on our smaller dataset of citations manually-annotated with citation purpose (described in Section \ref{subsec:cpa}). This helps to improve confidence in our findings. 

\begin{figure}
    \centering
    \includegraphics[width=\linewidth]{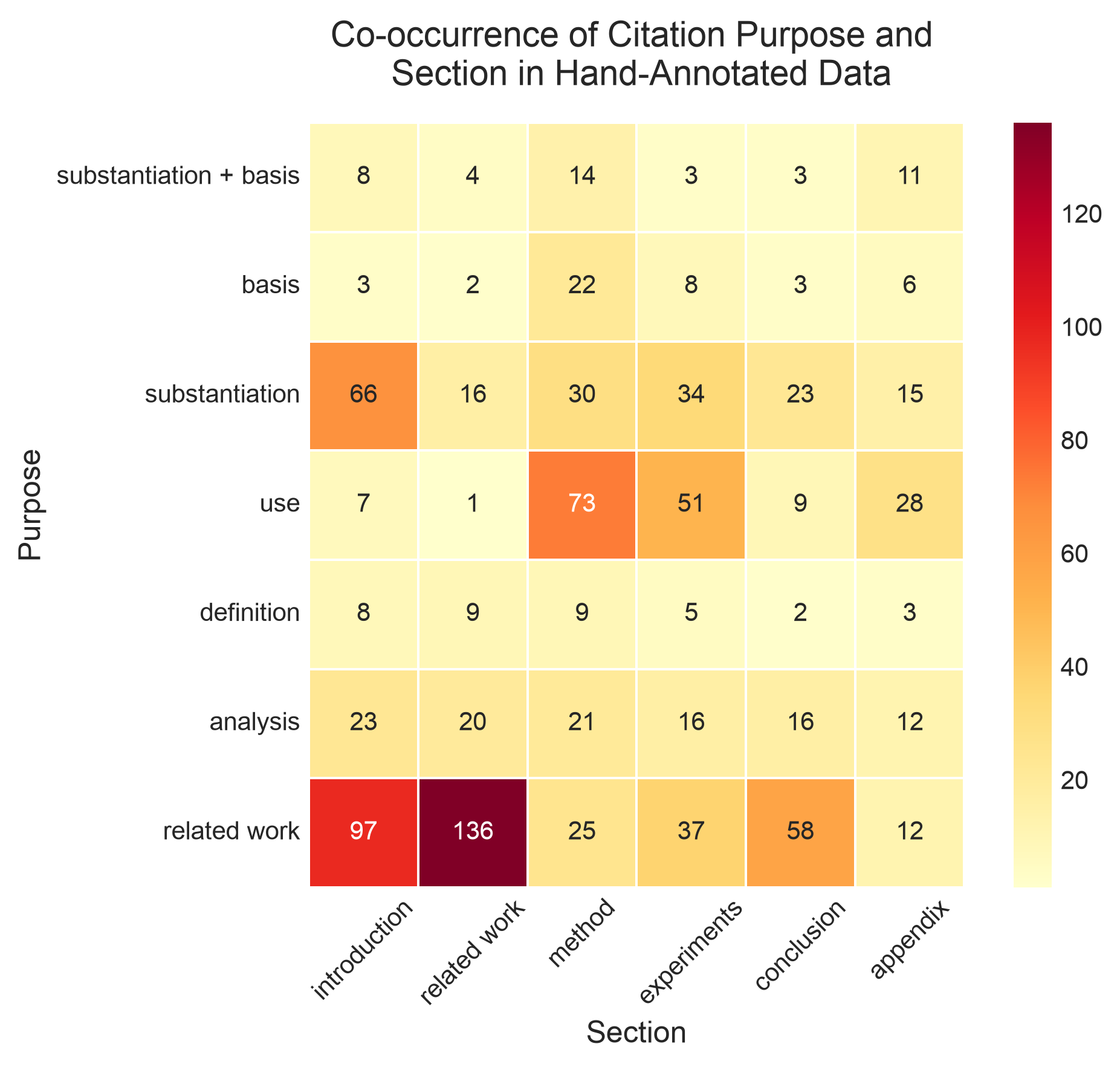}
    \caption{Co-occurrence of citation purpose and section in the annotated dataset reflects trends in the full dataset.}
    \label{fig:ann-sec-purp-cooccur}
\end{figure}

To further support our claim that sections and citation purposes intuitively co-occur, we replicate the trends shown in Figure \ref{fig:purpose-section-cooccurrence} on the hand-annotated purpose labels in Figure \ref{fig:ann-sec-purp-cooccur}. Citations with a \textit{Related Work} purpose label frequently appear in the \textit{Introduction} and \textit{Related Work} sections. \textit{Substantiation}-purpose citations occur in the \textit{Introduction} and \textit{Use}-purpose citations appear in the \textit{Method} and \textit{Experiments} sections. Some of the weaker trends found in the full dataset, such as a greater frequency of \textit{Analysis} citations in the \textit{Introduction} and \textit{Related Work} sections, are also found in the annotated data

\begin{figure}
    \centering
    \includegraphics[width=\linewidth]{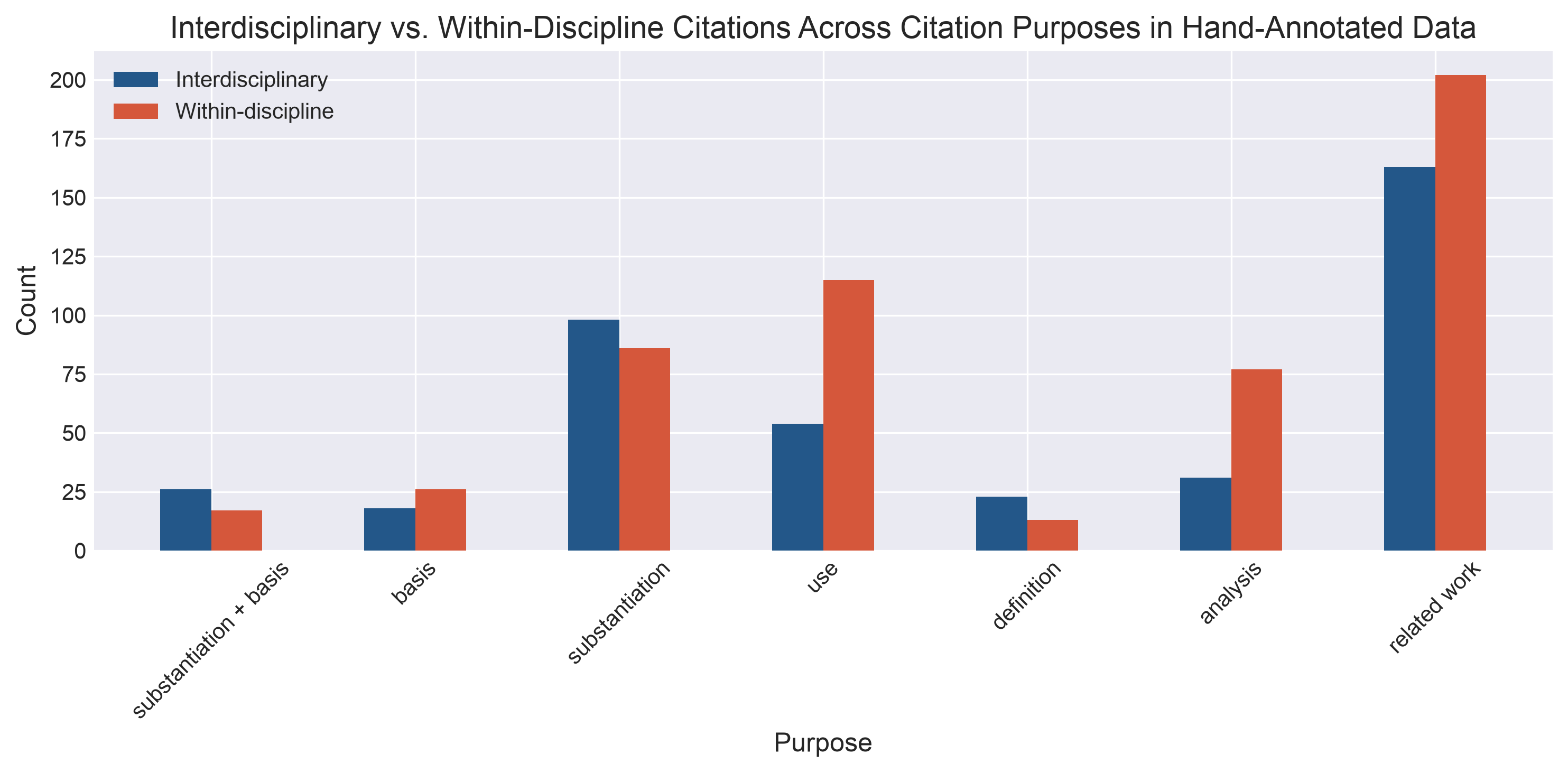}
    \caption{Interdisciplinary vs. Within-Discipline citations by citation purpose in hand-annotated dataset reflects trends in the full dataset.}
    \label{fig:ann-in-vs-out-eng}
\end{figure}

We also include Figure \ref{fig:ann-in-vs-out-eng} to show that our claim that the \textit{Substantiation} and \textit{Definition} citation purposes are more associated with interdisciplinary citations holds in the annotated dataset. 

\section{Error Analysis}
\label{app:error-analysis}

Here we outline potential failure points in our data collection and analysis framework.

\subsection{Failure to Match In-text References to Corresponding Citations} Typically, when you click on an in-text reference in an ACL paper, you are linked to the corresponding entry in the References section. Grobid captures this information, extracting a navigation id which allows us to match in-text references to corresponding citations. However, sometimes this navigation id is missing, either due to parsing errors, or because a paper author did not properly format an in-text reference. Because of this, not all in-text references are properly extracted and matched to citations.

\subsection{Failure to Match Citations to Field-of-Study} Sometimes citations in the \textit{References Section} do not contain enough metadata to match them to confidently match them to a field of study due to parsing errors or errors in citations. If we are unable to map citations to a field of study using the Semantic Scholar API or string-matching approach, we do not consider them in our analysis. 

\subsection{Incorrect Section Assignments} The string matching approach used to match section titles to canonical sections (detailed in Appendix \ref{app:sec-mapping}) can lead to incorrect assignments. Substring matching on section titles can be misleading. For example, the keyword \say{approach} is used to map to \textit{Method}. However, it is possible that an \textit{Experiments} section title would contain the substring, such as in \say{Our Approach vs. Prior Work}. 

\subsection{Incorrect Citation Purpose Assignments} Because we use an automatic classifier to predict Citation Purpose categories for our analysis in Section \ref{sec:case-study}, it carries some uncertainty. We detailed the performance of the classifier in Section \ref{subsec:cp-classifier} to acknowledge this.

\section{Codebook}
\label{app:codebook}

Our Citation Purpose Codebook is included in Appendix \ref{app:codebook-tb}. We also include a pdf version along with our codebase and data.
\renewcommand{\thetextbox}{\thesection.\arabic{textbox}}
\begin{textbox*}[htbp]
\centering
\begin{tcolorbox}[
    colback=lightblue,            
    colframe=darkblue,            
    width=\textwidth,              
    arc=3mm,                       
    boxrule=1pt,                   
    left=5mm,                      
    right=5mm,                     
    top=3mm,                       
    bottom=3mm,                    
    title={\textbf{Citation Purpose Codebook}}      
]
{\ttfamily\small

The 7 categories below describe possible purposes for an out-of-discipline citation in an academic paper. Each category contains a description and guidelines for when the category is pertinent, an example, and an explanation of why that particular category is right for the example. \\

\paragraph{I. Use} The method or tool in the citation is used directly in the citing work, with no or very trivial modifications.\\ 

\textbf{Abstract:} In this work, we carry out a data archaeology to infer books that are known to ChatGPT and GPT-4 using a name cloze membership inference query. We find that OpenAI models have memorized a wide collection of copyrighted materials, and that the degree of memorization is tied to the frequency with which passages of those books appear on the web. The ability of these models to memorize an unknown set of books complicates assessments of measurement validity for cultural analytics by contaminating test data; we show that models perform much better on memorized books than on nonmemorized books for downstream tasks. We argue that this supports a case for open models whose training data is known.\\
\textbf{Section:} experiments\\
\textbf{Citation Context:} Our second experiment predicts the amount of time that has passed in the same 250-word passage sampled above, using the conceptual framework of \textbf{Underwood (2018)}, explored in the context of GPT-4 in Underwood (2023). Unlike the year prediction task above, which can potentially draw on encyclopedic knowledge that GPT-4 has memorized, this task requires reasoning directly about the information in the passage, as each passage has its own duration. Memorized information about a complete book could inform the prediction about all passages within it (e.g., through awareness of the general time scales present in a work, or the amount of dialogue within it), in addition to having exact knowledge of the passage itself. As above, our core question asks: do we see a difference in performance between books that GPT-4 has memorized and those that it has not? To assess this, we again identify the top and bottom decile of books by their GPT-4 name cloze accuracy. We manually annotate the amount of time passing in a 250-word passage from that book using the criteria outlined by \textbf{Underwood (2018)}, including the examples provided to GPT-4 in Underwood (2023), blinding annotators to the identity of the book and which experimental condition it belongs to. We then pass those same passages through GPT-4 using the prompt in figure 4, adapted from Underwood (2023). We calculate accuracy as the Spearman $\rho$ between the predicted times and true times and calculate a 95\% confidence interval over that measure using the bootstrap.\\

\textbf{The above citation refers to a framework that is used directly in the citing work. There is no explicit reference to modifying this framework, or building upon it, and so it is inferred that the framework is used as is, and thus categorized as ‘Use’ and not ‘Basis’.}
\\
\paragraph{II. Definition}
The citation is used to define a topic or phrase (but NOT validate it - see `Substantiation').\\

\textbf{Abstract:} This paper analyses two hitherto unstudied sites sharing state-backed disinformation, Reliable Recent News (rrn.world) and WarOnFakes (waronfakes.com), which publish content in Arabic, Chinese, English, French, German, and Spanish. We describe our content acquisition methodology and perform cross-site unsupervised topic clustering on the resulting multilingual dataset. We also perform linguistic and temporal analysis of the web page translations and topics over time, and investigate articles with false publication dates. We make publicly available this new dataset of 14,053 articles, annotated with each language version, and additional metadata such as links and images. The main contribution of this paper for the NLP community is in the novel dataset which enables studies of disinformation networks, and the training of NLP tools for disinformation detection.\\
\textbf{Section:} introduction\\
\textbf{Citation Context:} Propaganda is defined as content that intentionally influences opinion to advance its creators' goals \textbf{(Bolsover and Howard, 2017)}. Numerous propaganda datasets have previously been created, with both document-level (Rashkin et al., 2017;Barrón-Cedeño et al., 2019) and span-level (Da San Martino et al., 2019b) technique annotations, using articles collected from multiple disinformation sites. At article-level, classifiers using combinations of multiple linguistic representations based on style and readability outperform content representation (Barrón-Cedeño et al., 2019), whereas content-based transformer models such as BERT have seen use at span-level (Da San Martino et al., 2019a). Detectors are often evaluated on single datasets, prompting concerns on generalisation (Martino et al., 2020).\\

\textbf{The above citation offers a definition for “propaganda” that is used in the citing work.}
}
\end{tcolorbox}
\end{textbox*}

\renewcommand{\thetextbox}{\thesection.\arabic{textbox}}
\begin{textbox*}[htbp]
\centering
\begin{tcolorbox}[
    colback=lightblue,            
    colframe=darkblue,            
    width=\textwidth,              
    arc=3mm,                       
    boxrule=1pt,                   
    left=5mm,                      
    right=5mm,                     
    top=3mm,                       
    bottom=3mm                     
]
{\ttfamily\small

\paragraph{III. Substantiation} The citation is used to verify or substantiate theoretical, empirical or methodological claims made in the citing work. Claims that describe the state of the literature (e.g. "There exist studies on this subject", "Multiple works have adopted this approach") are NOT valid claims for this category, and these citations might do better under 'Examples' or 'Related Work.' It does NOT matter if the claim is central to the premise of the citing work (such as claims made about contributions, or any claims in the abstract), or if the claim is local to the citation context. This category applies in any case where the citation is being used as proof of the claim. If classifying as ‘Substantiation,’ it is useful to identify the claim that is being supported.\\

\textbf{Abstract:} Prior work has revealed that positive words occur more frequently than negative words in human expressions, which is typically attributed to positivity bias, a tendency for people to report positive views of reality. But what about the language used in negative reviews? Consistent with prior work, we show that English negative reviews tend to contain more positive words than negative words, using a variety of datasets. We reconcile this observation with prior findings on the pragmatics of negation, and show that negations are commonly associated with positive words in negative reviews. Furthermore, in negative reviews, the majority of sentences with positive words express negative opinions based on sentiment classifiers, indicating some form of negation.\\
\textbf{Section:} introduction\\
\textbf{Citation Context:} A battery of studies have validated the Pollyanna hypothesis that positive words occur more frequently than negative words in human expressions, using corpora ranging from Google Books to Twitter (Dodds et al., 2015;Garcia et al., 2012;Boucher and Osgood, 1969; \textbf{Kloumann et al., 2012}). The typical interpretation is connected with the positivity bias, which broadly denotes 1) a tendency for people to report positive views of reality, 2) a tendency to hold positive expectations, views, and memories, and 3) a tendency to favor positive information in reasoning (Carr, 2011; Augustine et al., 2011; Hoorens, 2014). However, it remains an open question whether the Pollyanna hypothesis holds in negative reviews, where the communicative goal is to express negative opinions.\\

\textbf{The citation offers proof of the claim “positive words occur more frequently than negative words in human expressions.” While the fact of this evidence is mentioned explicitly in the citation context above, this explicit mention is not necessary to apply this category to a citation.}\\

\paragraph{IV. Basis}
The method, idea, or tool used in the citation is used as a basis for the method, idea, or tool in the citing work. It must be explicitly stated that this is being built upon, or it must be very evident from the abstract provided that this is an idea that is central to the premise and is thus being explored further.\\

\textbf{Abstract:} Online misogyny is a pernicious social problem that risks making online platforms toxic and unwelcoming to women. We present a new hierarchical taxonomy for online misogyny, as well as an expert labelled dataset to enable automatic classification of misogynistic content. The dataset consists of 6,567 labels for Reddit posts and comments. As previous research has found untrained crowdsourced annotators struggle with identifying misogyny, we hired and trained annotators and provided them with robust annotation guidelines. We report baseline classification performance on the binary classification task, achieving accuracy of 0.93 and F1 of 0.43. The codebook and datasets are made freely available for future researchers.\\
\textbf{Section:} method\\
\textbf{Citation Context:} This taxonomy draws on the typologies of abuse presented by Waseem et al. (2017) and \textbf{Vidgen et al. (2019)} as well as theoretical work in online misogyny research (Filipovic, 2007;Mantilla, 2013;Jane, 2016;Ging, 2017;Anzovino et al., 2018;Ging and Siapera, 2019;Farrell et al., 2019). It was developed by reviewing existing literature on online misogyny and then iterating over small samples of the dataset. This deductive-inductive process allowed us to ensure that conceptually distinct varieties of abuse are separated and that different types of misogyny can be unpicked. This is important given that they can have very different impacts on victims, different causes, and reflect different outlooks and interests on the part of the speaker.\\

\textbf{The citation is classified as ‘Basis’ because the authors state that their taxonomy (central to their work, although this factor is not necessary) “draws on” the typology presented in the citation.}
}
\end{tcolorbox}
\end{textbox*}

\renewcommand{\thetextbox}{\thesection.\arabic{textbox}}
\begin{textbox*}[htbp]
\centering
\begin{tcolorbox}[
    colback=lightblue,            
    colframe=darkblue,            
    width=\textwidth,              
    arc=3mm,                       
    boxrule=1pt,                   
    left=5mm,                      
    right=5mm,                     
    top=3mm,                       
    bottom=3mm                     
]
{\ttfamily\small

\paragraph{V. Substantiation + Basis}
The citation is used as proof of a claim made in the citing work (see category III above), and this claim is either directly built upon in the citing work, or used as the basis for a specific idea (see category IV above).\\

\textbf{Abstract:} Multiple studies have demonstrated that behavior on internet-based social media platforms can be indicative of an individual's mental health status. The widespread availability of such data has spurred interest in mental health research from a computational lens. While previous research has raised concerns about possible biases in models produced from this data, no study has quantified how these biases actually manifest themselves with respect to different demographic groups, such as gender and racial/ethnic groups. Here, we analyze the fairness of depression classifiers trained on Twitter data with respect to gender and racial demographic groups. We find that model performance systematically differs for underrepresented groups and that these discrepancies cannot be fully explained by trivial data representation issues. Our study concludes with recommendations on how to avoid these biases in future research.\\
\textbf{Section:} method\\
\textbf{Citation Context:} Tokenization. Raw text within in Tweets was tokenized using a modified version of the Twokenizer (O'Connor et al., 2010). English contractions were expanded, while specific retweet tokens, username mentions, URLs, and numeric values were replaced by generic tokens. As pronoun usage tends to differ in individuals living with depression \textbf{(Vedula and Parthasarathy, 2017)}, we removed any English pronouns from our stop word set (English Stop Words from nltk.org). Case was standardized across all tokens, with a single flag included if an entire post was made in uppercase letters.\\

\textbf{The citation offers proof for the claim “pronoun usage tends to differ in individuals living with depression,” and this claim is used to justify their removal of English pronouns in their approach.}\\

\paragraph{VI. Analysis}
The method or idea in the citation is being analyzed in some way by presenting comparisons or critiques. These analyses must have a direct link with the citing work, as evidenced in the citation context or the abstract. For example, the citation’s idea or method may be compared to one in the citing work, or may be criticized and its limitations used to justify an approach taken in the citing work. There must be no explicit mention of building on the citation’s idea in the citing work.\\

\textbf{Abstract:} Natural language inference (NLI) is the task of determining whether a piece of text is entailed, contradicted by or unrelated to another piece of text. In this paper, we investigate how to tease systematic inferences (i.e., items for which people agree on the NLI label) apart from disagreement items (i.e., items which lead to different annotations), which most prior work has overlooked. To distinguish systematic inferences from disagreement items, we propose Artificial Annotators (AAs) to simulate the uncertainty in the annotation process by capturing the modes in annotations. Results on the CommitmentBank, a corpus of naturally occurring discourses in English, confirm that our approach performs statistically significantly better than all baselines. We further show that AAs learn linguistic patterns and context-dependent reasoning.\\
\textbf{Section:} method\\
\textbf{Citation Context:} If we view the AAs as a committee of three members, our architecture is reminiscent of the Query by Committee (QBC) \textbf{(Seung et al., 1992)}, an effective approach for active learning paradigm. The essence of QBC is to select unlabeled data for labeling on which disagreement among committee members (i.e., learners pre-trained on the same labeled data) occurs. The selected data will be labeled by an oracle (e.g., domain experts) and then used to further train the learners. Likewise, in our approach, each AA votes for an item independently. However, the purpose is to detect disagreements instead of using disagreements as a measure to select items for further annotations. Moreover, in our AAs, the three members are trained on three disjoint annotation partitions for each item (see Section 3.2).\\

\textbf{The citation refers to an architecture that is similar to the one proposed in the citing work. It is clear from the citation context that the authors of the citing work are contrasting their own architecture with the one in the citation.}\\
}
\end{tcolorbox}
\end{textbox*}

\renewcommand{\thetextbox}{\thesection.\arabic{cb_textbox}}
\begin{cb_textbox*}[htbp]
    \centering
    \begin{tcolorbox}[
        colback=lightblue,            
        colframe=darkblue,            
        width=\textwidth,              
        arc=3mm,                       
        boxrule=1pt,                   
        left=5mm,                      
        right=5mm,                     
        top=3mm,                       
        bottom=3mm                     
    ]
    {\ttfamily\small
    
    \paragraph{VII. Related Work}
    The citation refers to work that is related to the citing work or provides it as an example of a method, idea, or phenomenon discussed without further description. It does not fit any of the other categories.\\
    
    \textbf{Abstract:} Methods and applications are inextricably linked in science, and in particular in the domain of text-as-data. In this paper, we examine one such text-as-data application, an established economic index that measures economic policy uncertainty from keyword occurrences in news. This index, which is shown to correlate with firm investment, employment, and excess market returns, has had substantive impact in both the private sector and academia. Yet, as we revisit and extend the original authors' annotations and text measurements we find interesting text-as-data methodological research questions: (1) Are annotator disagreements a reflection of ambiguity in language? (2) Do alternative text measurements correlate with one another and with measures of external predictive validity? We find for this application (1) some annotator disagreements of economic policy uncertainty can be attributed to ambiguity in language, and (2) switching measurements from keyword-matching to supervised machine learning classifiers results in low correlation, a concerning implication for the validity of the index.\\
    \textbf{Section:} experiments\\
    \textbf{Citation Context:} KeyExp. Although Baker et al. use human auditors to find policy keywords that minimize the false positive and false negative rates, they do not expand or optimize for economy or uncertainty keywords. Thus, we expand these keyword lists via GloVe word embeddings 9 (Pennington et al., 2014), and find the five nearest neighbors via cosine distance. 10 This is a simple keyword expansion technique. In future work, one could look to the literature on lexicon induction to improve creating lexicons that represent the semantic concepts of interest (Taboada et al., 2011;Pryzant et al., 2018;\textbf{Hamilton et al., 2016};Rao and Ravichandran, 2009). Alternatively, one could also create a probabilistic classifier over pre-selected lexicons to soften the predictions, or use other uncertainty lexicons or even automatic uncertainty cue detectors.\\
    
    \textbf{As inferred from the given context, the citation describes work related to creating lexicons that represent the semantic concepts of interest, which is identifiable as related to the citing work.  This is classified as ‘Related Work,’ and not ‘Examples,’ because there are multiple ideas being discussed, and it is not clear which, if any, the citation is supposed to be an example of. Additionally, the ideas being referenced are potential future directions, implying a certain level of novelty; it is therefore unlikely that the citation contains a concrete example of the specified idea.}

    }
    \end{tcolorbox}
    \caption{Codebook for Citation Purpose annotations.}
    \label{app:codebook-tb}
\end{cb_textbox*}

\section{Prompts}
\label{app:prompts}
The prompt used for Citation Purpose prediction are included in Appendix \ref{prompt:cp_class} and \ref{prompt:cp_class_desc}.
\renewcommand{\thetextbox}{\thesection.\arabic{textbox}}
\begin{textbox*}[htbp]
\centering
\begin{tcolorbox}[
    colback=lightgreen,            
    colframe=darkgreen,            
    width=\textwidth,              
    arc=3mm,                       
    boxrule=1pt,                   
    left=5mm,                      
    right=5mm,                     
    top=3mm,                       
    bottom=3mm                     
]
{\ttfamily\small

\textbf{System Prompt}\\
\\
You are an expert in academic citation analysis. Your task is to classify the purpose of a citation in a scientific paper.\\
\\
\textbf{User Prompt}\\
\\
Given the following citation context and citation text, choose the most appropriate category from the options below. Each option also has two examples to help you understand the category. Please read the category descriptions and examples carefully before making your choice.\\
\\
Citation context: <citation context>\\
\\
Citation text: <citation text>\\
\\
Candidate categories: <categories>\\
\\
Respond with ONLY the category name in a JSON object, exactly as written above. Do not include any explanation.\\
\\
Here is the expected output format: \{``category'': ``{category name}''\}
}
\end{tcolorbox}
\caption{Predicting Citation Purpose.}
\label{prompt:cp_class}
\end{textbox*}

\renewcommand{\thetextbox}{\thesection.\arabic{textbox}}
\begin{textbox*}[htbp]
\centering
\begin{tcolorbox}[
    colback=lightgreen,
    colframe=darkgreen,
    width=\textwidth,
    arc=3mm,
    boxrule=1pt,
    left=5mm,
    right=5mm,
    top=3mm,
    bottom=3mm
]
{\ttfamily\small

\textbf{substantiation + basis}
\begin{enumerate}[leftmargin=*, nosep, topsep=0pt]
    \item The citation implicitly or explicitly provides evidence for a theoretical, empirical, or methodological claim.
    \item The above is true AND the claim does NOT describe the state of the literature (e.g.\ ``There exist studies on this subject'', ``Multiple works have adopted this approach'').
    \item The above are true AND the claim is NOT being used as the basis for the citing work's idea or method.
    \item The above are true AND the citation includes a specific method or idea that is being used, built upon, improved, or otherwise modified.
    \item The above are true AND the citation is NOT used directly.
\end{enumerate}
\vspace{4mm}

\textbf{substantiation}
\begin{enumerate}[leftmargin=*, nosep, topsep=0pt]
    \item The citation implicitly or explicitly provides evidence for a theoretical, empirical, or methodological claim.
    \item The above is true AND the claim does NOT describe the state of the literature (e.g.\ ``There exist studies on this subject'', ``Multiple works have adopted this approach'').
    \item The above are true AND the claim is NOT being used as the basis for the citing work's idea or method.
\end{enumerate}
\vspace{4mm}

\textbf{basis}
\begin{enumerate}[leftmargin=*, nosep, topsep=0pt]
    \item The citation refers to an idea being used as a basis by the citing work.
    \item The citation refers to a method that is explicitly being built upon, improved, or modified by the citing work. It is NOT used directly.
    \item There is no explicit mention but it is very obvious from the abstract that this is integral to the citing work's idea and is thus being built upon.
    \item One of the above is true AND there is NO theoretical or empirical claim associated with the idea or method that motivates its use.
\end{enumerate}
\vspace{4mm}

\textbf{definition}
\begin{enumerate}[leftmargin=*, nosep, topsep=0pt]
    \item The citation helps define or explain a topic or phrase.
    \item The above is true AND there is no justification, validation, or proof otherwise provided.
\end{enumerate}
\vspace{4mm}

\textbf{analysis}
\begin{enumerate}[leftmargin=*, nosep, topsep=0pt]
    \item The context offers a critique of the citation's work.
    \item The citation's work is explicitly compared to the citing work.
    \item The citation's work is being used as a baseline to compare with the citing work's approach.
    \item One of the above is true, with NO explicit mention of building upon, using, or modifying the citation's work.
\end{enumerate}
\vspace{4mm}

\textbf{use}
\begin{enumerate}[leftmargin=*, nosep, topsep=0pt]
    \item The citation refers to a specific concrete method, not an idea, being used by the citing work.
    \item The above is true AND there is no explicit mention of building upon, improving, or modifying the citation's method in a non-trivial way.
\end{enumerate}
\vspace{4mm}

\textbf{related work}
\begin{enumerate}[leftmargin=*, nosep, topsep=0pt]
    \item The citation is related to the article.
    \item The above is true AND the citation does not fall under any of the other categories.
\end{enumerate}

}
\end{tcolorbox}
\caption{Category descriptions included in prompt.}
\label{prompt:cp_class_desc}
\end{textbox*}

\end{document}